\documentclass[10pt,journal]{IEEEtran}
\usepackage{amsmath}
\usepackage{amssymb}
\usepackage{algorithm}
\usepackage{algorithmicx}
\usepackage{algpseudocode}
\usepackage{graphicx}
\usepackage{subfigure}
\usepackage{epstopdf}
\usepackage{cite}
\usepackage{textcomp}
\usepackage{mathrsfs}
\usepackage{color}
\usepackage{booktabs}
\usepackage{cases}
\usepackage{setspace}
\usepackage{bm}
\usepackage{cuted}
\usepackage{booktabs}
\usepackage{stfloats}
\usepackage{makecell}
\usepackage{mathtools}
\usepackage{amssymb}
\usepackage{pdfrender}
\usepackage{bbding}
\begin{document}
	
	\title{Exploiting Constructive Interference for Backscatter Communication Systems} 	
	
	\author{Bowen~Gu, \emph{Student Member, IEEE},  Dong~Li, \emph{Senior Member, IEEE}, \\Ye~Liu, 	\emph{Member, IEEE}, Yongjun~Xu, \emph{Senior Member, IEEE} 
		\IEEEcompsocitemizethanks{
			\IEEEcompsocthanksitem This work was supported in part by the Science and Technology Development Fund, Macau, SAR, under Grant 0018/2019/AMJ, Grant 0110/2020/A3, and Grant 0029/2021/AGJ,  in part by National Natural Science Foundation of China (62271094, U21A20448),  in part by Key Fund of Natural Science Foundation of Chongqing (CSTB2022NSCQ-LZX0009),  and in part by the Scientific and Technological Research Program of Chongqing Municipal Education Commission (KJZD-K202200601). (\textit{Corresponding author: Dong Li}.)
			\IEEEcompsocthanksitem Bowen Gu and Dong Li are with the School of Computer Science and Engineering, Macau University of Science and Technology, Avenida Wai Long, Taipa, Macau 999078, China (e-mails: gubwww@163.com, dli@must.edu.mo).
			\IEEEcompsocthanksitem Ye Liu is  with the School of Computer Science and Engineering, Macau University of Science and Technology, Avenida Wai Long, Taipa, Macau 999078, China, and  also with the College of Artificial Intelligence, Nanjing Agricultural University, Nanjing 210031, China (e-mail: yeliu@njau.edu.cn).
			\IEEEcompsocthanksitem Yongjun Xu is with the School of Communication and Information Engineering, Chongqing University of Posts and Telecommunications, Chongqing 400065, China (e-mail: xuyj@cqupt.edu.cn).
		}	
}

	\maketitle
	\thispagestyle{empty}
	\pagestyle{empty}
	
	\begin{abstract}
	Backscatter communication (BackCom), one of the core technologies to realize zero-power communication, is expected to be a pivotal paradigm for the next generation of the Internet of Things (IoT).  However, the ``strong" direct link (DL) interference (DLI) is traditionally assumed to be harmful, and generally drowns out the ``weak" backscattered signals accordingly, thus deteriorating the performance of BackCom. In contrast to the previous efforts to eliminate the DLI, in this paper,  we exploit the constructive interference (CI), in which the DLI contributes to the backscattered signal.  To be specific, our objective is to maximize  the received signal-to-noise ratio (SNR) by jointly optimizing the receive beamforming vectors and tag selection factors under different detection error probability (DEP) requirements, which leads to two different optimization problems. However, the resulting problems are non-convex and unanalyzable due to constraints on the DEP. To solve these problems, the Kullback-Leibler divergence is first applied to transform the DEP into a tractable form. Then, inspired by the alternating optimization, we respectively propose two successive convex approximation (SCA)-based algorithms to solve the corresponding sub-problems with beamforming design, and a greedy algorithm to solve the sub-problem with tag selection. In order to gain insight into the CI, we consider a special case with the single-antenna reader to reveal the channel angle between the backscattering link (BL) and the DL, in which the DLI will become constructive.  Simulation results show that significant performance gain can always be achieved with the proposed algorithms compared to the traditional algorithms without the CI in terms of the received SNR. The derived constructive channel angle for the BackCom system with a single-antenna reader is also confirmed by simulation results.
	\end{abstract}
	\IEEEpeerreviewmaketitle
	\section{Introduction}	
	
	As the number of connected devices has grown by leaps and bounds over the past few decades, the Internet of Things (IoT) is transforming industry, logistics, and everyday use in an unforeseen manner, ushering in a new industrial revolution \cite{VaseziCST2022,XuGuiGacaninAdachi}. Recent market studies envision that the total number of connected IoT devices will reach 83 billion by 2024, rising from 35 billion connections in 2020 \cite{CentenaroCST2021}. However, this is a mixed blessing, as further promotion of the IoT will necessitate a huge expense to maintain the operation of these ever-increasing devices. Recently, backscatter communication (BackCom) has been emerging  as a promising technology that can break the gridlock of these power-hungry devices, which are replaced by passive tags without the need for active radio frequency (RF) components \cite{VanHuynhCST2018}. In the BackCom system, the externally generated carrier by a dedicated power source is utilized to carry the tag’s signal, and only the impedance loading is applied to backscatter the received signal, which can significantly reduce the hardware cost and pave the way for zero-power communication \cite{ToroWuLeungtgcn2022}. 

    However, it should be noted that, in the BackCom system, the backscattered signal actually undergoes double-channel fading, which is to blame for the quite weak signal strength at the reader. It is challenging to recover the backscattered data from different tags due to the mutual interference caused by the concurrent transmission \cite{JinHeMengFang2021}. On the other hand, another issue ensues: when the signal from the direct link (DL) encounters these weak signals at the reader, the resulting interference is strong enough to drown out the desired signal, which significantly degrades the performance of the BackCom system, especially the signal detection and transmission rate \cite{BharadiaJoshi2015}. Therefore, it is hard to overstate the importance of interference suppression for BackComs.
    
     \begin{table*}[t]
    	\centering
    	\caption{Summary of the Related Literature with Interference Processing in BackComs}
    	\begin{tabular}{|c|c|c|c|c|c|c|c|}
    		\hline
    		\multicolumn{3}{|c|}{Summary for interference processing}& \multicolumn{2}{c|}{Interference is constructive or not?}\\ 
    		\hline
    		Ref. &Main processing Method&Processing result&Mutual Interference&DLI \\
    		\hline
    		[9]& SIC &Interference cancellation & {\XSolidBrush}&{\XSolidBrush} \\
    		\hline
    		[11]-[14] & FS & Interference avoidance &{\XSolidBrush}&{\XSolidBrush} \\
    		\hline
    		[15]-[17] & OFDM & Interference avoidance &{\XSolidBrush}&{\XSolidBrush} \\
    		\hline
    		[18]-[20] & TDMA & Interference avoidance & {\XSolidBrush}&{\XSolidBrush} \\
    		\hline
    		[21]-[23] & Tag selection & Interference avoidance & {\XSolidBrush}&{\XSolidBrush} \\
    		\hline
    		[24]-[25] & Phase offset & Interference utilization & {\CheckmarkBold} &{\XSolidBrush} \\
    		\hline
    		[26] & SLP & Interference utilization & {\CheckmarkBold} &{\XSolidBrush} \\
    		\hline
    		Our work& KLD-based method & Interference utilization & {\XSolidBrush} & {\CheckmarkBold}\\
    		\hline
    	\end{tabular}
    \end{table*}

    In fact, interference suppression has been fruitfully studied in BackCom systems. One of the most typical methods is the successive interference cancellation (SIC) technique \cite{PatelHoltzman1994,XuQinGuiGacainSari}, by which the strong interference signal in the mixed signal can be stripped.  However, it should be noted that the SIC may suffer from residual interference, and  interference cancellation for multiple weak signals may not perform well. In this regard, the frequency shift (FS) technique has been proposed as a promising method to mitigate the interference in BackCom systems \cite{HuZhangRostamisigcom2016}. To be specific, each backscatter tag was able to shift its carrier signal to adjacent non-overlapping frequency bands and isolate  the backscattered signal from the DL signal so that the mutual interference between different tags and the DL interference (DLI) could be avoided \cite{ZhangRostamiHusigcom2016,ZhangJosBha2017,LiLiangtec2019,LIwcl2019}.  Furthermore,  BackCom systems over orthogonal frequency division multiplexing (OFDM) carriers were proposed in \cite{YangLiangglobe2016,Darsenaaccess2019,Darsenageill}, which canceled out the DLI by exploiting the repeating structure of the ambient OFDM signals with the aid of the cyclic prefix. Besides, time division multiple access (TDMA) (see, \cite{XuGuLi2021,GuLiXu,Litcomtwobirds}) and tag selection (see, \cite{LIPengHutvt2018,XuGuHuLi,ZhangGaoFan2019}) have been also widely used in BackCom systems to avoid the mutual interference among different tags.
    
    \subsection{Motivations and Contributions}
    
 It is noted that, most of the existing works assume that the DLI is always harmful, and thus focus on how to avoid the effect of interference for BackCom systems. However, there are several natural questions that arise: Is interference necessarily destructive in all cases?  Is it possible to blaze a trail to turn the destructive interference into a constructive one? These problems have received little attention, and there have been only a few research efforts paid to answer the above questions. The constructive interference (CI) was achieved among multiple cooperative tags in centralized scenarios \cite{DobkinFreedGerdom2015} and decentralized scenarios \cite{PerezHermansVoigt2017}, respectively, in which several tags backscatter the same packet synchronously, which makes it possible that backscattered signals can be constructively combined, leading to an increased signal strength at the reader. Besides,  a BackCom system with the symbol level precoding (SLP) technique was proposed in \cite{ZhengWenChen2022} to convert the multi-user interference into the constructive region of the desired data symbol by designing the transmit beamforming on the symbol level.  
    
Although the CI was considered in \cite{DobkinFreedGerdom2015} and \cite{PerezHermansVoigt2017} among different tags in  BackCom systems, which means that CI can be found among multiple weak signals, they ignore the real protagonist, namely, the strong interference from the DL, since compared with the mutual interference among tags, the DLI is more harmful. In \cite{ZhengWenChen2022}, the multi-user interference in broadcasting channels was considered to be converted into the CI, where the tag signal detection was not taken into account and there was actually no DLI. 
In this regard, we take up the baton in this paper to probe how to convert the DLI into a helpful interference in enhancing the backscattered signal strength, which is in contrast to most of the existing works that aim at canceling the DLI \cite{XuQinGuiGacainSari,HuZhangRostamisigcom2016,ZhangRostamiHusigcom2016,ZhangJosBha2017,LiLiangtec2019,LIwcl2019,YangLiangglobe2016,Darsenaaccess2019,Darsenageill,XuGuLi2021,LIPengHutvt2018,Litcomtwobirds,GuLiXu,XuGuHuLi}. Besides, for more visualization, Table I briefly summarizes the differences between this paper and the mentioned works. To the best of our knowledge, this is the first attempt to harness the DLI for the backscattered signal for BackCom systems. Specifically, we consider a multi-tag BackCom system with a dedicated power source (PS) supplying energy to the tags, and a multi-antenna reader to enhance the reception efficiency.

   In a nutshell, the contributions of this paper are summarized as follows:
    \begin{itemize}
  	\item To evaluate whether the DLI is constructive or not, the detection error probabilities (DEPs) under the cases with and without the DL are evaluated. To make our analysis tractable, a Kullback-Leibler divergence (KLD)-oriented approach is proposed, where the KLDs under these two cases are  requisitioned. By doing so, the CI can be obtained.  However, the mutual interference among multiple tags also hinders the exploitation of the CI, and we consider tag selection to avoid this problem. It is worth pointing out that the KLD analysis of the DLI is new, and not presently available in existing works. 
    	\item Our objective is to maximize the received  signal-to-noise ratio (SNR) under the two different CI requirements (i.e., the consensual and evolved CIs) with the single- and multi-antenna PSs, respectively, via beamforming design and tag selection, which leads to two different optimization problems in each scenario. Moreover, the formulated problems are non-convex and hard to solve. In order to circumvent this difficulty, we decompose these problems based on the alternating optimization method. Then, the corresponding sub-problems for beamforming design are respectively solved by proposing two different  algorithms with successive convex approximation (SCA),  and the sub-problems for tag selection are solved via greedy search. 
    	\item In order to gain more insight into the CI, we consider a special case with one single antenna at the reader. In this case, we are able to derive the closed-form expressions for the feasible region of the input SNR and the channel angle, where the evolved CI exists. This result is new and unique in the sense of its potential to build up the CI in practice.
    	\item Simulation results show that significant performance gain can always be achieved with the proposed algorithms compared with the benchmark algorithms in terms of the received SNR. Besides, the CI region is demonstrated for the case with multiple antennas at the reader, and the derived angle for the case with single antenna at the reader is also confirmed by simulation results. Moreover, we also demonstrate from the derived angle that it is more likely to build up the CI for the relatively large DEP tolerance or large path loss.
    \end{itemize}

 \subsection{Organization}
    The remainder of the paper is organized as follows:
    In Section II, we present the system model and analyze the DEPs for two different cases. In Section III, we formulate two optimization problems, and two different greedy-based algorithms with SCA are proposed to solve them. Besides, a special case is also considered for more insight in Section III. Section IV gives the simulation results, and Section V concludes this paper.

  \subsection{Notations}
 Throughout the paper, scalars, vectors, and matrices are denoted by lowercase, boldface lowercase, and boldface uppercase letters, respectively. $|\cdot|$ and $||\cdot||$ denote the absolute value of a complex scalar and the $l_2$-norm of a vector, respectively. $(\cdot)^H$, det$(\cdot)$, Rank$(\cdot)$,  Tr$(\cdot)$, $||\cdot||_*$, $||\cdot||_{1}$, and $||\cdot||_{2}$ represent the Hermitian, the determinant, the rank,  the trace, the nuclear norm, the $\mathcal{L}_1$ norm, and the spectral norm of the matrix, respectively. $\mathcal{CN}(\mu,\sigma^2)$ is the circularly symmetric complex Gaussian (CSCG) distribution with mean $\mu$ and variance $\sigma^2$. $\lambda^{\max}(\cdot)$ and $\lambda^{\min}(\cdot)$ denote the maximum and minimum eigenvalues of the matrix, respectively. $\bm{I}_N$ represents the $N\times N$ identity matrix. $W_0(\cdot)$ and $W_{-1}(\cdot)$ denote  the principal branch and the bottom branch of the Lambert $W$ function, respectively.

	\section{System Model And Detection Analysis}

	\subsection{System Architecture }
	We consider a BackCom system with single input multiple output (SIMO), as shown in Fig. 1, with a single-antenna PS, $K$ single-antenna tags ($k\in\mathcal{K}=\{1,2,\cdots,K\}$), and a reader with $M$ antennas ($m\in\mathcal{M}=\{1,2,\cdots,M\}$). In particular, all tags are powered by the PS so that each tag can backscatter its information to the reader. The channel coefficients of the PS-tag $k$ link, the PS-reader link, and the tag $k$-reader link are respectively given by $h_{\text{ST}}^k\in \mathbb{C}^{1\times1}$, $\bm h_{\text{SR}}\in \mathbb{C}^{M\times1}$, and $\bm h_{\text{TR}}^k\in \mathbb{C}^{M\times1}$. Assume that channels between the transmitter and the receiver follow a block-fading based model, namely, $h_{\text{ST}}^k$, $\bm h_{\text{SR}}$, and $\bm h_{\text{TR}}^k$ remain constant during backscattering in one time interval, but vary independently in different intervals. It is further assumed that the PS transmits $N$ CSCG signals $s(n)$, $n\in\mathcal{N}=\{1,2,\cdots,N\}$ with the transmitted power $\sigma_s^2$ in one time interval \cite{LiuWeiNgYuan}. Moreover, $s(n)$ for $ n = \{1,2, \cdots,N\} $ are independent and identically distributed. The signal received at the $k$-th tag can be expressed as
		\begin{align} \label{ar1}
		r_k(n)= h_{\text{ST}}^{k} s(n)+ z_k(n),
	\end{align}
	where $z_k(n)$ denotes the noise at $k$-th tag.  Then, the $k$-th tag backscatters the signal $r_k(n)$ and transmits its own binary signal $B_k$.  Note that the noise backscattered by each tag can be ignored in the backscattering process due to its poor signal strength. Consequently, the backscattered  signal by the $k$-th tag can be expressed as 
	\begin{align} \label{ar2}
		c_k(n)=  \alpha_k B_k h_{\text{ST}}^{k} s(n),
	\end{align}
where  $\alpha_k \in [0,1]$  represents the attenuation inside the tag, which is achieved by adjusting the load impedance at the port of the antenna, and also depends on the structure mode of the antenna \cite{LuNiyatoJiang}.

	    \begin{figure}[t]
	\centerline{\includegraphics[width=2.8in]{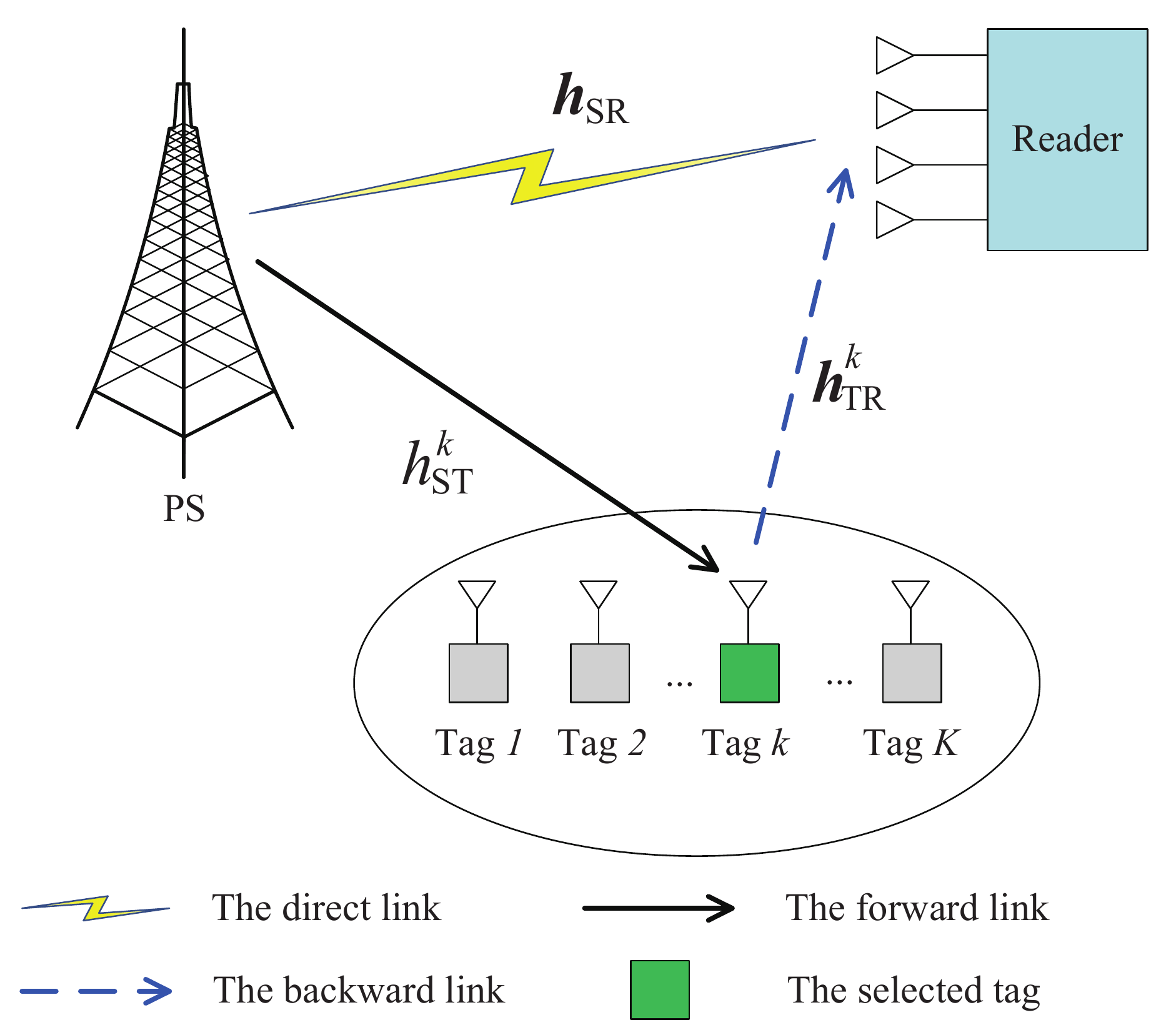}}
	\caption{A multi-tag SIMO-based BackCom system.}
	\label{fig1}
\end{figure}
	
In order to avoid the mutual interference among tags, only one tag can be selected to backscatter its information at one slot. Noting that the signal received by the reader is not only from the selected tag (e.g., the $k$-th tag), but also from the PS. Thus, the signal received by the reader can be expressed as 
	\begin{align} \label{0}
		\bm y_k(n)=\underbrace{\bm h_{\text{SR}} s(n)}_{\text{Direct signal}}+\underbrace{ \bm h_{\text{TR}}^{k} c_{k}(n)}_{\text{Backscattering signal}}+\underbrace{ \bm w(n)}_{\text{Noise}},
	\end{align}
 where $\bm w(n)\in \mathbb{C}^{M\times1}$ denotes the noise vector of the reader and is assumed to be an i.i.d. CSCG random vector with $\bm w(n)\sim\mathcal{CN}(0,\sigma_w^2\bm I_M)$.  $\sigma_w^2$ denotes the noise power.  
	By using the receive beamforming, the received signal at the reader can be expressed as
	\begin{equation}
		\begin{aligned} \label{1}
			y_k(n)&{=}\bm v_k^H\bm h_{\text{SR}} s(n){+} \bm v_k^H\bm h_{\text{TR}}^{k}c_{k}(n){+}{\bm v_k^H \bm w(n)},
		\end{aligned}	
	\end{equation}
	where $\bm v_k\in \mathbb{C}^{M\times1}$ denotes the receive beamforming vector at the reader.
	
Note that this framework of multi-tag selection for the Backcom system is beneficial for interference-free transmission and can efficiently combat the
double-fading channel, which can be well used in radio frequency identification (RFID) \cite{ZhangGaoFanJin} for logistics management, wireless body area networks (WBANs) \cite{LingHuLiHan} for health monitoring and data sensing, etc.
	
	\subsection{CSI Assumption}
 It is assumed that the BackCom system operates in the time-division multiplexing (TDD) mode, and thus the channel state information (CSI) can be obtained by the training-based channel estimation at the uplink transmission \cite{MishraLarsson}.  Specifically, the reader is designed to transmit two sets of pilots in two training phases \cite{LongLiangGuoYang}.  In the first phase, the $k$-th tag switches its impedance into an initial matched state, and then it receives the pilot signals from the reader as excitation and does not backscatter any signal.  Thus, the PS can estimate the CSI for the DL (i.e., $\bm h_{\text{SR}}$) via channel reciprocity.  In the second phase, the $k$-th tag switches its impedance into a fixed and known backscatter state to represent a symbol $B_k^0$ after receiving the training signals in the first phase.  Thus, the PS can estimate the composite channel $\bm h_{\text{SR}}+B_k^0h_{\text{ST}}^k\bm h_{\text{TR}}^k$ via the training pilots.  Then, for the given $B_k^0$, the CSI of the backscattering link (BL) (i.e., $h_{\text{ST}}^k\bm h_{\text{TR}}^k$) can be obtained at the PS by subtracting the estimated the DL channel component from the estimated composite channel.
	
\subsection{Signal Model}
	
	In this subsection, we give some analysis for the received signal when the DLI is available or not. Specifically, when the DL is available, the received signal of the reader can be rewritten with a binary hypothesis. In particular, the null hypothesis ($\mathcal H_0$) is that  the $k$-tag does not backscatter the incident signal and the alternative hypothesis ($\mathcal H_1$) is that the $k$-tag backscatters the incident signal. In this regard, we have
	\begin{equation} \label{2} 	
		y_k(n)\text{=}\left\{ \begin{aligned}
			& \mathcal H_0:\bm v_k^H\bm h_0s(n)+\bm v_k^H\bm w(n),~B_k=0, \\ 
			& \mathcal H_1:\bm v_k^H\bm h_1^ks(n)+\bm v_k^H\bm w(n),~B_k=1,\\ 
		\end{aligned} \right.
	\end{equation}
	where $\bm h_0=\bm h_{\text{SR}}$, $\bm h_1^k=\bm h_{\text{SR}}+\bm h_{\text{STR}}^k$, and $\bm h_{\text{STR}}^k=\alpha_kh_{\text{ST}}^k\bm h_{
		\text{TR}}^k$.
	
	Thus, the distributions of the received signal $y_k(n)$ can be expressed as 
	\begin{equation} \label{3} 	
		\left\{ \begin{aligned}
			&\mathcal H_0:y_k(n)\sim\mathcal{CN}( 0, \delta_0^k),~B_k=0, \\ 
			&\mathcal H_1:y_k(n)\sim\mathcal{CN}( 0, \delta_1^k),~B_k=1,\\ 
		\end{aligned} \right.
	\end{equation}
where the variances under $\mathcal H_0$ and $\mathcal H_1$ can be expressed as
	\begin{equation} \label{4} 	
	\delta_i^k=	\left\{ \begin{aligned}
			&  |\bm v_k^H\bm h_0|^2\sigma_s^2+\sigma_w^2||\bm v_k||^2,i=0, \\
			&  |\bm v_k^H\bm h_1^k|^2\sigma_s^2+\sigma_w^2||\bm v_k||^2, i=1.\\
		\end{aligned} \right.
	\end{equation}
	
	 Accordingly, the received signal $\bm{y}_k$ at the reader in one time interval obeys the following distributions
	\begin{equation} \label{4a} 	
		\left\{ \begin{aligned}
			&\mathcal H_0:\bm{y}_k\sim\mathcal{CN}(\bm 0,\textbf{C}_0^k),~ B_k=0, \\ 
			&\mathcal H_1:\bm{y}_k\sim\mathcal{CN}(\bm 0,\textbf{C}_1^k),~ B_k=1,\\ 
		\end{aligned} \right.
	\end{equation}
	where the  covariance matrices under $\mathcal H_0$ and $\mathcal H_1$ can be expressed as $\textbf{C}_i^k= \delta_i^k\bm I_N, i\in\{0,1\}.$

	Let $p_0(\bm{y}_k)$ and $p_1(\bm{y}_k)$ denote the likelihood functions of the received signal of the reader under $\mathcal H_0$ and $\mathcal H_1$, respectively. Based on (\ref{4a}), we have
	\begin{equation}\label{5}
		\begin{aligned} 
			p_i(\bm{y}_k)&=\frac{1}{\pi^N\det (\textbf{C}_i^k)}\exp(-\bm{y}_k^{H}({\textbf{C}_i^k})^{-1}\bm{y}_k)\\
			&=\frac{1}{\pi^N (\delta_i^k)^N}\exp(-\frac{\bm{y}_k^{H}\bm{y}_k}{\delta_i^k}), i\in\{0,1\}.\\
		\end{aligned}
	\end{equation}

	As  a counterpart, assuming the DL is unavailable, the received signal at the reader can be expressed as
	\begin{equation}
		\begin{aligned} \label{11a}
			\bar y_k(n)= \bm {v}_k^H\bm h_{\text{TR}}^{k}c_{k}(n){+}{\bm { v}_k^H \bm w(n)}.
		\end{aligned}	
	\end{equation}

	Accordingly, the received signal $ \bm{\bar y}_k $ in one time interval obeys the following distributions
		\begin{equation} \label{11} 	
		\left\{ \begin{aligned}
			&\mathcal H_0:\bm{\bar y}_k\sim\mathcal{CN}(\bm 0,\bar {\textbf{C}}_0^k),~B_k=0, \\ 
			&\mathcal H_1:\bm{\bar y}_k\sim\mathcal{CN}(\bm 0,\bar {\textbf{C}}_1^k),~B_k=1,\\ 
		\end{aligned} \right.
	\end{equation}
	where $	\bar {\textbf{C}}_i^k=\bar \delta_i^k\bm I_N$ are the covariance matrices, and 
\begin{equation} \label{10b} 	
	\bar \delta_i^k=	\left\{ \begin{aligned}
		& |\bm { v}_k^H\bm {\bar h}_0|^2\sigma_s^2+\sigma_w^2||\bm { v}_k||^2,i=0,\\
		& |\bm {v}_k^H\bm {\bar h}_1^k|^2\sigma_s^2+\sigma_w^2||\bm { v}_k||^2,i=1,\\
	\end{aligned} \right.
\end{equation}
where $\bm {\bar h}_0=\bm 0$, and $\bm {\bar h}_1^k=\bm h_{\text{STR}}^k$.

In this case, the likelihood functions of the received signal of the reader under $\mathcal H_0$ and $\mathcal H_1$, respectively, can be expressed as 
	\begin{equation}\label{5a}
	\begin{aligned} 
		\bar p_i(\bm{\bar y}_k)=\frac{1}{\pi^N (\bar \delta_i^k)^N}\exp(-\frac{\bm{\bar y}_k^{H}\bm{\bar y}_k}{\bar \delta_i^k}), i\in\{0,1\}.\\
	\end{aligned}
\end{equation}

\subsection{Detection Probability}

In BackCom systems, the goal for the reader is to distinguish two different hypotheses $\mathcal H_0$ or $\mathcal H_1$ by applying a specific decision rule. In general, the priori probabilities of hypotheses $\mathcal H_0$ and $\mathcal H_1$ are assumed to be equal. As such, the detection error probability (DEP) of the reader is defined as in \cite{MaZhangLu,YanCongHanlyZhou}
	\begin{equation}\label{5b}
	\begin{aligned} 
\xi={\text{Pr}}(\mathcal{D}_1|{\mathcal H}_0)+{\text{Pr}}(\mathcal{D}_0|{\mathcal H}_1),
	\end{aligned}
\end{equation}
where $\mathcal D_1$ and $\mathcal D_0$ are the binary decisions that infer whether backscattering  transmission is present or not, respectively. ${\text{Pr}}(\mathcal{D}_1|{\mathcal H}_0)$ and ${\text{Pr}}(\mathcal{D}_0|{\mathcal H}_1)$ denote the false alarm and miss detection probabilities, respectively. 

Taking the case with the DL as an example, for an optimal detector at the reader, the optimal  DEP ($\xi_k^*$) for detecting the backscattered signal by the $k$-th tag can be obtained as  \cite{YanCongHanlyZhou,ZhangXiao}
	\begin{equation}\label{5c}
	\begin{aligned} 
		  \xi_k^*=1-\mathcal{V}_{T}(p_0, p_1),
	\end{aligned}
\end{equation}
where $\mathcal{V}_{T}(p_0, p_1)$ denotes  the total variation between $p_0(\bm{y}_k)$ and $p_1(\bm{y}_k)$. 

Similarly, the optimal DEP ($\zeta_k^*$) of the reader for detecting the backscattered signal by the $k$-th tag under the case without the DL can be obtained as 
	\begin{equation}\label{5ca}
	\begin{aligned} 
		\zeta_k^*=1-\mathcal{V}_{T}(\bar p_0, 	\bar p_1),
	\end{aligned}
\end{equation}
where $\mathcal{V}_{T}(\bar p_0, 	\bar p_1)$ denotes  the total variation between $\bar p_0(\bm{\bar y}_k)$ and $\bar p_1(\bm{\bar y}_k)$.

\section{Problem Formulation and Algorithm Design}

 In this section, we first formulate two optimization problems for maximizing the received SNR with different CI requirements under a single-antenna PS, and propose two iterative algorithms with SCA  to solve them after some designed transformation. Then, the different problems for received SNR are also studied by extending to a multi-antenna PS. Besides, a closed-form expression for the upper bound on the angle between the DL and the BL is revealed to shed light on the CI for the single-antenna reader.
 
  \subsection{Problem Formulation for  Consensual CI}
 
 To evaluate the enhancement of the BackCom transmission with the assistance of the DL,  an optimization problem for maximizing the received SNR at the reader is  formulated by the joint optimization for the receive beamforming design and the tag selection, which is given by
 \begin{equation} \label{p1} 
 	\begin{aligned}	
 		& \underset{\bm v_k, \beta_k}{\mathop{\max}}\, \sum\limits_{k=1}^K \beta_k |\bm v_k^H\bm h_1^k|^2 \gamma \\ 
 		&\text{s.t.}~
 		{{C}_{1}}:\sum\limits_{k=1}^{K}{\beta_k}=1, \beta_k\in \{0,1\},\\ 
 		&\quad ~~{{C}_{2}}:||\bm v_k||^2=1,\forall k,\\ 	
 		&\quad ~~{{C}_{3}}:  \sum\limits_{k=1}^K \beta_k \xi_k^{*}\le  \xi^{\max},\\ 
 		&\quad ~~{{C}_{4}}:  \sum\limits_{k=1}^K \beta_k \zeta_k^{*}\le \zeta^{\max},\\ 
 	\end{aligned}
 \end{equation}  
 where  $\beta_k$ is the tag selection factor, i.e., $\beta_k=1$ denotes the $k$-th tag is selected, otherwise $\beta_k=0$. $\gamma=\frac{\sigma_s^2}{\sigma_w^2}$ denotes the input SNR.  $\xi^{\max}$ and $\zeta^{\max}$ denote the tolerated threshold of the DEPs with and without the DL for the BackCom system.
 For problem (\ref{p1}), $C_1$ is the tag selection constraint. $C_2$ denotes the receive beamforming constraint. $C_3$ and $C_4$ are the corresponding DEP constraints, which guarantee that the DEPs with and without the DL are not greater than the corresponding tolerable DEPs, respectively. It is noted that, by setting $\xi^{\max}\le \zeta^{\max}$, the constructive DLI can be obtained, which is defined as \textit{the  consensual CI}.
 
 As can be seen, problem (\ref{p1}) is challenging  to solve directly due to the mixed-integer objective and the non-smooth constraints. To make it more treatable, assuming the $k$-th tag has been selected,  the subproblem for beamforming design can be transformed as 
 \begin{equation} \label{p1c} 
 	\begin{split}	
 		& \underset{\bm v_k}{\mathop{\max}}\,|\bm v_k^H\bm h_1^k|^2\gamma\\ 
 		&\text{s.t.}~C_2, {{C}_{3-1}}:  \xi_k^{*}\le  \xi^{\max},\\ 
 		&\quad ~~{{C}_{4-1}}:  \zeta_k^{*}\le \zeta^{\max}.\\ 
 	\end{split}
 \end{equation}

 \subsection{DEP Transformation}

  Unfortunately, there are still many obstacles in solving  problem (\ref{p1c}), one of which is that the total variation metrics in $C_{3-1}$ and $C_{4-1}$ are unwieldy and intractable for products of probability measures \cite{BashGoeckelTowsley}. To deal with that,  the Bretagnolle–Huber bound \cite{BHbound} is adopted, i.e., 
 \begin{equation}\label{5d}
 	\begin{aligned} 
 		\mathcal{V}_{T}(p_0, p_1)\le \sqrt{1-\exp(-D_{\text{KL}}^k(p_0||p_1))},
 	\end{aligned}
 \end{equation}
 or 
 \begin{equation}\label{5e}
 	\begin{aligned} 
 		\mathcal{V}_{T}(p_0, 	p_1)\le \sqrt{1-\exp(-D_{\text{KL}}^k(p_1||p_0))},
 	\end{aligned}
 \end{equation}
 where $D_{\text{KL}}^k(p_0||p_1)$ and $D_{\text{KL}}^k(p_1||p_0)$ denote the KLD from $p_1(\bm{y}_k)$ to $p_0(\bm{y}_k)$ and the KLD from $p_0(\bm{y}_k)$ to $p_1(\bm{y}_k)$, respectively. Specifically,  the expression of $D_{\text{KL}}^k(p_0||p_1)$ can be written as\footnote{It is noted that the Gaussian-distributed signal input is restrictively assumed in this paper, which is commonly used for analysis of communication systems due to its mathematical tractability. The impact of practical modulation type and order on the exploration of constructive interference will be another work, and is left for future work.}
 \begin{equation} \label{12}
 	\begin{aligned}
 		D_{\text{KL}}^k(p_0||p_1)&=\int_{\bm{y}_k}p_0(\bm{y}_k)\ln \frac{p_0(\bm{y}_k)}{p_1(\bm{y}_k)}d\bm{y}_k \\
 		&=N\ln \frac{\delta_1^k}{\delta_0^k}+N\frac{\delta_0^k}{\delta_1^k}-N.\\
 	\end{aligned}
 \end{equation}

Based on (\ref{5c}) and (\ref{5d}), we have the lower bound related to the KLD for $\xi_k^*$  under the case with the DL, which satisfies
 \begin{equation}\label{12a}
 	\begin{aligned} 
 		\xi_k^*\ge  1-\sqrt{1-\exp(-D_{\text{KL}}^k(p_0||p_1))}\triangleq \bar \xi_k.
 	\end{aligned}
 \end{equation}

 Similarly,  the KLD of the corresponding likelihood functions  under different hypotheses without the DL can be given by
 \begin{equation} \label{7}
 	\begin{aligned}
 		D_{\text{KL}}^k(\bar p_0||\bar p_1)=N\ln \frac{\bar \delta_1^k}{\bar \delta_0^k}+N\frac{\bar \delta_0^k}{\bar \delta_1^k}-N.\\
 	\end{aligned}
 \end{equation}
 
 Correspondingly,  the lower bound of the DEP related to the KLD  in the case without the DL can be expressed as 
 \begin{equation}\label{7a}
 	\begin{aligned} 
 		\zeta_k^*\ge  1-\sqrt{1-\exp (-D_{\text{KL}}^k(\bar p_0||\bar p_1))}\triangleq \bar \zeta_k.
 	\end{aligned}
 \end{equation}

Note that the maximum value of problem (\ref{p1}) can be achieved when $\xi_k^*$ and $\zeta_k^*$ are equal to their lower bounds, respectively, in (\ref{12a}) and (\ref{7a}).
Thus, $C_{3-1}$ and $C_{4-1}$ can be  rewritten by, respectively, 
  \begin{equation} \label{c1} 	
 	\left\{ \begin{aligned}
 		& \bar \xi_k\le  \xi^{\max},\\ 
        & \bar \zeta_k\le \zeta^{\max}, \\ 
 	\end{aligned} \right.
 \end{equation}	
  or equivalently
  \begin{equation} \label{c2a} 	
	\left\{ \begin{aligned}
		&D_{\text{KL}}^k(p_0||p_1)\ge-\ln(1-(1-\xi_k^{\max})^2\triangleq D^{\min},\\ 
		&D_{\text{KL}}^k(\bar p_0||\bar p_1)\ge -\ln(1-(1-\zeta^{\max})^2\triangleq 
		E^{\min}.\\ 
	\end{aligned} \right.
\end{equation}

Thus, problem (\ref{p1c}) can be transformed into the following problem, i.e., 
 \begin{equation} \label{p1b} 
	\begin{aligned}	
		& \underset{\bm v_k}{\mathop{\max}}\,  |\bm v_k^H\bm h_1^k|^2 \gamma \\ 
		&\text{s.t.}~C_2, C_{3-2}:D_{\text{KL}}^k(p_0||p_1)\ge D^{\min},\\ 
		&\quad ~~C_{4-2}: D_{\text{KL}}^k(\bar p_0||\bar p_1)\ge E^{\min}.\\ 
	\end{aligned}
\end{equation}

\subsection{Beamforming Design for Consensual CI}

However, it is difficult to solve problem (\ref{p1b}) due to the tight coupled relationship in $C_{3-2}$ and $C_{4-2}$. To deal with it,  we propose Theorem 1.

	\textit{\textbf{Theorem 1}}:  $C_{3-2}$ and $C_{4-2}$ can be transformed into the following equivalent forms, i.e.,	
	  \begin{equation} \label{tm1} 	
		\left\{ \begin{aligned}
			& \frac{(|\bm v_k^H\bm h_1^k|^2{-}|\bm v_k^H\bm h_0|^2)\gamma}{|\bm v_k^H\bm h_0|^2\gamma+1} \ge F(f^{\min})-1,\\
			&\gamma{|\bm {v}_k^H\bm {\bar h}_1^k|^2} \ge F(g^{\min})-1,\\
		\end{aligned} \right.
	\end{equation}
where $F(x)=\exp(W_{0}(-\exp(-x))+x)$, $f^{\min}\triangleq\frac{D^{\min}}{N}+1$, and $g^{\min}\triangleq\frac{E^{\min}}{N}+1$.

\textit{\textbf{Proof}}: Please see Appendix A.

Based on Theorem 1, problem (\ref{p1b}) can be transformed into the following problem, i.e.,
\begin{equation} \label{p1d} 
	\begin{split}	
		& \underset{\bm v_k}{\mathop{\max}}\,|\bm v_k^H\bm h_1^k|^2\gamma\\ 
		&\text{s.t.}~C_2, C_{3-3}: {(|\bm v_k^H\bm h_1^k|^2{-}|\bm v_k^H\bm h_0|^2)\gamma}\ge (F(f^{\min}){-}1)\\
		&\quad \quad \quad \quad ~~~~~~\cdot({|\bm v_k^H\bm h_0|^2\gamma+1}),\\
		&\quad ~~C_{4-3}:\gamma{|\bm {v}_k^H\bm {\bar h}_1^k|^2} \ge F(g^{\min})-1. \\
	\end{split}
\end{equation}
However, it is still hard to solve problem (\ref{p1d}) directly due to the non-convexity of $C_{3-3}$. To deal with this, the SCA-based method is applied \cite{WangYangMengZhan}.  In each iteration, we construct a global underestimator of the intractable term via its first-order Taylor approximation, which is given by
 \begin{equation}\label{sca1}
	\begin{aligned} 
|\bm v_k^H\bm h_1^k|^2{\ge} 2(\bm h_1^k(\bm h_1^k)^H\bm {\bar v}_k{(l)})^H\bm v_k{-}|\bm {\bar v}_k^H(l)\bm h_1^k|^2{\triangleq} \mu_k^{\text{BF}}(l),\\
	\end{aligned}
\end{equation} 
where $\bm {\bar v}_k(l)$ is the feasible point of $\bm {v}_k$ in the $l$-th iteration of the SCA.

As a result, problem (\ref{p1d}) can be finally expressed as 
\begin{equation} \label{p1e} 
	\begin{split}	
		& \underset{\bm v_k}{\mathop{\max}}\,\mu_k^{\text{BF}}(l)\gamma\\ 
		&\text{s.t.}~C_2, C_{4-3},\\
		&\quad ~~C_{3-4}^a: {(\mu_k^{\text{BF}}(l){-}|\bm v_k^H\bm h_0|^2)\gamma}\ge (F(f^{\min})-1)\\
		&\quad \quad \quad \quad ~~\cdot ({|\bm v_k^H\bm h_0|^2\gamma+1}),\\
		&\quad ~~C_{3-4}^b: |\bm v_k^H\bm h_1^k|^2\ge \mu_k^{\text{BF}}(l).\\
	\end{split}
\end{equation}
Note that problem (\ref{p1e}) is a standard convex problem, which can be solved by CVX.  The detail of this algorithm is shown in \textbf{Algorithm 1}. Here, the computational complexity  for Algorithm 1 is evaluated.  Given the number of
constraints, interior point methods with a vector variable $\bm v_k$ of size $M\times 1$ will take $\mathcal{O}(\sqrt{M} \ln (1/\epsilon))$ iterations, with each iteration requiring $\mathcal{O}(M^3)$ arithmetic operations for the worst case  \cite{LongLiangGuo}, where $\epsilon$ denotes the  precision of the interior point algorithm. Letting the maximum iteration number for SCA  is $L$,  the total computation complexity for beamforming design is $\mathcal{O}(LM^{3.5}\ln(1/\epsilon))$.

 \subsection{Problem Formulation for Evolved CI}
 
So far, we have studied the consensual CI for the BackCom system in the previous subsection. However, the relationship between $\xi_k^*$ and $\zeta_k^*$ is unknown. In this subsection, we are going to investigate another CI, i.e., evolved CI, by exploiting the following relationships between $\xi_k^*$ and $\zeta_k^*$, i.e.,  
	 \begin{equation} \label{c3a} 	
		\left\{ \begin{aligned}
			&\xi_k^*>\zeta_k^*, \forall k, \\ 
			&\xi_k^*\le \zeta_k^*, \forall k.\\ 
		\end{aligned} \right.
	\end{equation}		
It is worth mentioning that when $\xi_k^*>\zeta_k^*$ holds, the dilemma will be that the involvement of the DL leads to a larger DEP than the case without the DL, which would be intolerable for perfectionists if the DL is required to have no negative impact on the BackCom system.
On the contrary, when $\xi_k^*\le \zeta_k^*$ holds, we can assert that introducing the DL does not destroy the detection performance compared with the case without the DL, and can even improve it.

Motivated by this observation,  a new optimization problem for achieving the evolved CI can be reformulated, which is expressed  as
 \begin{equation} \label{p2a} 
	\begin{aligned}	
		& \underset{ \bm {\hat v}_k, \hat \beta_k}{\mathop{\max}}\,  \sum\limits_{k=1}^{K} \hat \beta_k |\bm {\hat v}_k^H\bm h_1^k|^2 \gamma \\ 
		&\text{s.t.}~
		\bar C_1:\sum\limits_{k=1}^{K}{\hat \beta_k}=1, \hat \beta_k\in \{0,1\},\\ 
		&\quad ~~\bar C_2: ||\bm {\hat v}_k||^2=1, \forall k, \\
		&\quad ~~\bar C_{3}: \sum\limits_{k=1}^{K}{\hat \beta_k} \xi_k^*\le \sum\limits_{k=1}^{K}{\hat \beta_k} \zeta_k^*,\\ 
		&\quad ~~\bar C_{4}: \sum\limits_{k=1}^{K}{\hat \beta_k} \zeta_k^{*}\le \zeta^{\max}.\\ 
	\end{aligned}
\end{equation}

	\begin{table}[t]
	\centering
	\begin{tabular}{>{\raggedleft}p{0.5cm}p{7.5cm}}
		\toprule
		\multicolumn{2}{l}{\textbf{Algorithm 1} The Algotithm  for Problem (\ref{p1e})}\\
		\toprule
		\multicolumn{2}{l} {\textbf{Input:} $K$, $M$, $N$, $\alpha_k$, $h^k_{\text{ST}}$, $\bm h_{\text{SR}}$, $\bm h^k_{\text{TR}}$, $\sigma_w^2$, $\sigma_s^2$, $ \xi^{\max}$, $\zeta^{\max}$.}\\
		\multicolumn{2}{l} {\textbf{Set:} Set the iteration index $l=1$, the tolerance $\omega$, the  }\\
		\multicolumn{2}{l} {\quad \quad maximum iteration number $L$. } \\
		\multicolumn{2}{l} {\textbf{Initialize:} Set initial point $\bm {\bar v}_k(1)$.} \\
		\multicolumn{2}{l} {\textbf{Output:} $\bm {v}_k^*$.} \\
		1: &\textbf{while} $|\mu_k^{\text{BF}}(l+1)-\mu_k^{\text{BF}}(l)|\ge \omega$ or $l \le L$ \textbf{do} \\
		2: &~~~~Given $\bm {\bar v}_k(l)$, solve problem (\ref{p1e}) by using CVX  \\
		&\quad  ~~to obtain $\mu_k^{\text{BF}}(l)$ and $\bm {v}_k(l)$.\\
		3: &~~~~Update $l=l+1$.\\
		4:& \textbf{end while}\\
		5:&Obtain the solution to problem (\ref{p1e}) as $\bm {v}_k^*=\bm {v}_k(l)$.\\
		\toprule
	\end{tabular}
\end{table}	

Assuming that the $k$-th tag has been selected, problem (\ref{p2a}) can be further rewritten as 
 \begin{equation} \label{p2ad} 
	\begin{aligned}	
		& \underset{ \bm {\hat v}_k}{\mathop{\max}}\,  |\bm {\hat v}_k^H\bm h_1^k|^2 \gamma \\ 
		&\text{s.t.}~\bar C_2, \bar C_{3-1}: \xi_k^*\le \zeta_k^*,\\ 
		&\quad ~~\bar C_{4-1}: \zeta_k^{*}\le \zeta^{\max}.\\ 
	\end{aligned}
\end{equation}

Based on the same approach for the DEP-to-KLD transformation, serving the purpose of maximizing the objective of problem (\ref{p2ad}), the left-hand items in $\bar C_{3-1}$ and $\bar C_{4-1}$ are replaced by their lower bounds, i.e., 
 \begin{equation} \label{p2ba} 
	 	\left\{ \begin{aligned}
		& \bar \xi_k\le  \zeta_k^*, \\ 
		& \bar \zeta_k\le \zeta^{\max}.  \\ 
	\end{aligned} \right.
\end{equation}

Further, to guarantee the CI evolution, the lower-bound relaxation is applied, thus (\ref{p2ba}) is transformed into 
 \begin{equation} \label{p2bb} 
	\left\{ \begin{aligned}
		& \bar \xi_k\le  \bar \zeta_k, \\ 
		& \bar \zeta_k\le \zeta^{\max}.\\ 
	\end{aligned} \right.
\end{equation}

Correspondingly, problem (\ref{p2ad}) can be further rewritten as 
 \begin{equation} \label{p2b} 
	\begin{aligned}	
		& \underset{\bm {\hat v}_k}{\mathop{\max}}\, |\bm {\hat v}_k^H\bm h_1^k|^2 \gamma \\ 
		&\text{s.t.}~\bar C_2,\bar C_{3-2}: \Delta D_{\text{KL}}^k\ge 0,\\ 
		&\quad ~~\bar C_{4-2}:D_{\text{KL}}^k(\bar p_0||\bar p_1)\ge E^{\min},\\ 
	\end{aligned}
\end{equation}
where 
\begin{equation} \label{c5}
	\begin{aligned}
		\Delta D_{\text{KL}}^k&=D_{\text{KL}}^k(p_0||p_1)-D_{\text{KL}}^k(\bar p_0||\bar p_1)\\		 
		&=N\left[\ln\frac{ \delta_1^k}{ \delta_0^k}{+}\frac{\delta_0^k}{\delta_1^k}-\left(\ln \frac{\bar \delta_1^k}{\bar \delta_0^k}{+}\frac{\bar \delta_0^k}{\bar \delta_1^k}\right)\right].\\
	\end{aligned}
\end{equation}

\begin{figure}[t]
	\centerline{\includegraphics[width=2.7in]{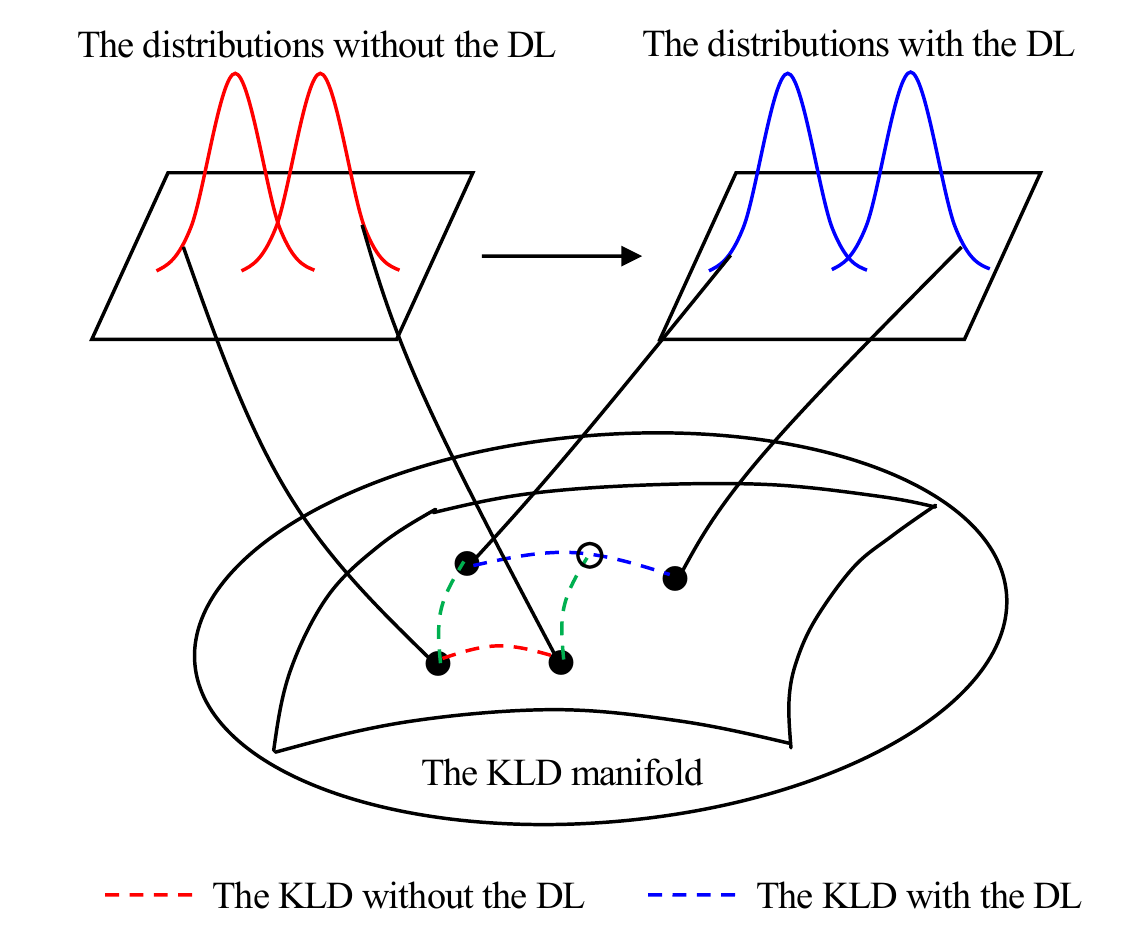}}
	\caption{A schematic manifold for the KLD.}
	\label{fig1a}
\end{figure}

\textbf{\textit{Remark 1:}}  By comparing problems (\ref{p1b}) and (\ref{p2b}),  it can be seen that the potential adverse condition that lurks in problem (\ref{p1b}) against the construction of CI  is removed in problem (\ref{p2b}). The purpose of problem (\ref{p2b}) is to explore whether it is always possible to create a larger KLD by using the DL compared with that without the DL, which is different from the former.   For illustration, Fig. \ref{fig1a} is provided to portray the relationship between different KLDs.  By doing so, we can effectively  enhance the backscattering strength without affecting or even reducing the DEP compared with the conventional BackCom system.

Note that problem (\ref{p2b}) can be regarded as evolving from  problem (\ref{p1b}), but the proposed algorithm for problem (\ref{p1b}) can not suitable for problem (\ref{p2b}) since $\bar C_{3-2}$ poses different challenge in solving the receive beamforming. This motivates us to design a new algorithm for problem (\ref{p2b}). 

	\subsection{Beamforming Design for Evolved CI}

In order to make problem (\ref{p2b}) more tractable,  we propose Theorem 2.
	
	\textit{\textbf{Theorem 2}}:  $\bar C_{3-2}$ can be transformed into the equivalent form, i.e.,
	\begin{equation} \label{m2a} 
		\begin{aligned}
			&{{|\bm {\hat v}_k^H\bm h_1^k|^2{-}|\bm {\hat v}_k^H\bm h_0|^2}{-}|\bm {\hat v}_k^H\bm {\bar h}_1^k|^2}\ge \gamma {|\bm {\hat v}_k^H\bm h_0|^2|\bm {\hat v}_k^H\bm {\bar h}_1^k|^2}.\\ 
		\end{aligned}
	\end{equation}

\textit{\textbf{Proof}}: Please see Appendix B.

Based on Theorem 1 and Theorem 2, problem (\ref{p2b}) can be transformed into the following problem, i.e.,
	\begin{equation} \label{17} 
		\begin{split}	
			& \underset{\bm {\hat v}_k}{\mathop{\max}}\,|\bm {\hat v}_k^H\bm h_1^k|^2\gamma\\ 
			&\text{s.t.}~\bar C_{2},  \bar C_{3-3}: {{|\bm {\hat v}_k^H\bm h_1^k|^2{-}|\bm {\hat v}_k^H\bm h_0|^2}{-}|\bm {\hat v}_k^H\bm {\bar h}_1^k|^2}\\
			&\quad \quad \quad \quad \quad ~~~  \ge \gamma {|\bm {\hat v}_k^H\bm h_0|^2|\bm {\hat v}_k^H\bm {\bar h}_1^k|^2},\\ 
			&\quad ~~ \bar C_{4-3}: \gamma{|\bm {\hat v}_k^H\bm {\bar h}_1^k|^2} \ge F(g^{\min})-1. \\
		\end{split}
	\end{equation}

Then, by introducing  a new variable $\bm W_k=\bm {\hat v}_k\bm {\hat v}_k^H$, $ \bar C_{3-3}$ and $\bar C_{4-3}$ can be transformed, respectively,  into
	\begin{equation} \label{17d} 
		\begin{split}	
	&\text {Tr}(\bm H_1^k\bm W_k){-}\text {Tr}(\bm H_0\bm W_k){-}\text {Tr}(\bm {\bar H}_1^k\bm W_k)\\
	&\quad \quad \quad \quad ~~~{\ge}\gamma  \text {Tr}(\bm H_0\bm W_k)\text {Tr}(\bm {\bar H}_1^k\bm W_k),\\
		\end{split}
	\end{equation}
and
	\begin{equation} \label{17e} 
	\begin{split}	
		\gamma {\text {Tr}(\bm {\bar H}_1^k\bm W_k)} \ge F(g^{\min})-1,\\
	\end{split}
\end{equation}
	where $\bm H_1^k=\bm h_1^k(\bm h_1^k)^H$, $\bm H_0=\bm h_0\bm h_0^H$, and $\bm {\bar H}_1^k=\bm {\bar h}_1^k(\bm {\bar h}_1^k)^H$.
	
	Thus, problem (\ref{17}) can be expressed as 
	\begin{equation} \label{16a} 
		\begin{split}	
			& \underset{\bm W_k}{\mathop{\max}}\,\text {Tr}(\bm H_1^k\bm W_k)\gamma\\ 
			& \text{s.t.}~
			\bar C_{2-1}: \text{Tr} (\bm W_k)=1,\\
			&\quad ~~\bar C_{3-4}:\text {Tr}(\bm H_1^k\bm W_k){-}\text {Tr}(\bm H_0\bm W_k){-}\text {Tr}(\bm {\bar H}_1^k\bm W_k)\\
			&\quad \quad \quad \quad  ~~ {\ge}\gamma  \text {Tr}(\bm H_0\bm W_k)\text {Tr}(\bm {\bar H}_1^k\bm W_k),\\ 
			&\quad ~~\bar C_{4-4}: \gamma {\text {Tr}(\bm {\bar H}_1^k\bm W_k)}\ge F(g^{\min})-1, \\
			&\quad ~~\bar C_5: \text{Rank}(\bm W_k)=1.
		\end{split}
	\end{equation}
	However,  problem (\ref{16a}) is still a non-convex problem caused by the non-convexity in $\bar C_{3-4}$. In general, it is difficult or even intractable to obtain the global optimal solution to such a non-convex problem. To make it more treatable,  we introduce  a slack variable $t_k$ such that $\text {Tr}(\bm H_0\bm W_k)\le t_k$, problem (\ref{16a}) can be transformed into the following problem, i.e.,
			\begin{equation} \label{18} 
			\begin{split}	
				& \underset{\bm W_k, t_k}{\mathop{\max}}\,\text {Tr}(\bm H_1^k\bm W_k)\gamma\\ 
				&\text {s.t.}~
				\bar C_{2-1},\bar C_{4-4}, \bar C_5,\\
				&\quad~~\bar C_{3-5}:\text {Tr}(\bm H_1^k\bm W_k){-}t_k{-}\text {Tr}(\bm {\bar H}_1^k\bm W_k)\\
				&\quad \quad \quad \quad~~  \ge\gamma  t_k\text {Tr}(\bm {\bar H}_1^k\bm W_k),\\
				&\quad~~\bar C_6:\text {Tr}(\bm H_0\bm W_k)\le t_k.\\
			\end{split}
		\end{equation}
It is noted that the rank-one constraint $\bar C_5$ is an impediment to solving problem (\ref{18}). Semi-definite relaxation (SDR) has been commonly adopted to tackle the rank-one constraint, which, however, may not result in a rank-one matrix. Moreover, some approximation methods, such as Gaussian randomization, may cause performance deterioration when the number of antennas is quite large  \cite{XuYuSunNg}. To deal with this, $\bar C_5$ is equivalently transformed into the following difference-of-convex function \cite{YangJiangShiDing}, such as 
	\begin{equation} \label{sp3} 	
	\begin{aligned}
	||\bm W_{k}||_*-||\bm W_{k}||_{2}=0,
	\end{aligned}
\end{equation}
where $||\bm W_k||_*=\sum\limits_{i}\delta_i(\bm W_k)$ and $||\bm W_k||_{2}=\delta_1(\bm W_k)$.  It is noted that for any $\bm X$ belonging to the positive semi-definite matrix, $||\bm X||_*\ge ||\bm X||_{2}=\underset{i}{\mathop{\max}}~\delta_i(\bm X)$ is always held. The inequality can be treated as an equation  if and only if $\bm X$ is a rank-one matrix,  especially, $\delta_1(\bm X)=||\bm X||_{2}$ and $\delta_i(\bm X)=0, \forall i=2,\cdots, M.$ As a result, (\ref{sp3}) can be efficiently met for the rank-one constraint $\bar C_5$. However, (\ref{sp3}) is still non-convex. To deal with it,  a penalty function is employed, and thus problem (\ref{18}) can be equivalently transformed into the following problem, such as 
	\begin{equation} \label{sp4} 
	\begin{split}	
		& \underset{\bm W_k, t_k}{\mathop{\max}}\,\text {Tr}(\bm H_1^k\bm W_k)\gamma -\chi (	||\bm W_k||_*-||\bm W_k||_{2})\\
		&\text{s.t.}~	\bar C_{2-1},\bar C_{3-5}, \bar C_{4-4}, \bar C_6.	
	\end{split}
\end{equation} 
where $\bar C_5$ is relaxed to a penalty term added to the objective function, and $\chi$ is a constant which penalizes the objective function for any matrix whose rank is not one.

However, problem (\ref{sp4}) is still an intractable problem with any $\chi>0$ due to the non-convexity of the objective function. To deal with this, the SCA method is applied. Specifically, for a given $\bm W_k(j)$ in the $j$-th iteration of SCA, a convex upper bound for the penalty term can be obtained by using first-order Taylor expansion, which can be expressed as
	\begin{equation} \label{sp5} 	
	\begin{aligned}
		||\bm W_k||_*-||\bm W_k||_{2}\le ||\bm W_k||_*-||\bar {\bm W}_{k}(j)||, \\
	\end{aligned}
\end{equation}
where $\bar {\bm W}_{k}(j)\triangleq ||\bm W_k(j)||_{2}+ {\rm Tr}\left[ \bar {\bm u}\left( \bm W_k(j)\right)  \left( \bar {\bm u}(\bm W_k(j))\right) ^{\rm H}\left( \bm W_k-\bm W_k(j)\right) \right] $ and $\bar {\bm u}\left( \bm W_k(j)\right) $ represents the eigenvector corresponding to the largest eigenvalue of $\bm W_k(j)$ \cite{YangJiangShiDing}.

	\begin{table}[t]
	\centering
	\begin{tabular}{>{\raggedleft}p{0.5cm}p{7.5cm}}
		\toprule
		\multicolumn{2}{l}{\textbf{Algorithm 2} The Algotithm  for Problem (\ref{sp6})}\\
		\toprule
		\multicolumn{2}{l} {\textbf{Input:} $K$, $M$, $N$, $\alpha_k$, $h^k_{\text{ST}}$, $\bm h_{\text{SR}}$, $\bm h^k_{\text{TR}}$, $\sigma_w^2$, $\sigma_s^2$, $ \xi^{\max}$, $\zeta^{\max}$.}\\
		\multicolumn{2}{l} {\textbf{Set:} Choose a large $T$. Define $\Delta{t_k}=(t_k^{\text{ub}}-t_k^{\text{lb}}) /T$.}\\
		\multicolumn{2}{l} {\textbf{Initialize:} Set initial point $\gamma^{\text{INI}}= 0$.} \\
		\multicolumn{2}{l} {\textbf{Output:} $\bm {\hat v}_k$.} \\
		1: &\textbf{for} $i=0:T$ \\
		2: &~~~~Set $t_k=i\Delta t_k+t_k^{\text{lb}}$.\\
		3: &~~~~Set initial point $\bm W_k(1)$. iteration index $j = 1$. \\
		4:&~~~~\textbf{repeat}\\
		5: &~~~~~~~With given $\bm W_k(j)$, obtain the intermediate \\
		&~~~~~~~solution $\bm W_k^*$ by solving problem (\ref{sp6}).\\
		6:&~~~~~~~Update $j=j+1$ and $\bm W_k^*=\bm W_k(j)$.\\
		7:&~~~~\textbf{until convergence} \\
		8: &~~~~\textbf{if} $ \text {Tr}(\bm H_1^k\bm W_k^*)\gamma \ge \gamma^{\text{INI}}$\\
		9:& ~~~~~~~~Update $\gamma^{\text{INI}}= \text {Tr}(\bm H_1^k\bm W_k^*)\gamma$.\\
		10: &~~~~\textbf{end if}\\
		11:& \textbf{end for}\\
		12:& Recover $\bm {\hat v}_k^*$ from $\bm W_k^*$.\\
		\toprule
	\end{tabular}
\end{table}	

Thus, problem (\ref{sp4}) can be finally rewritten as 
		\begin{equation} \label{sp6} 
		\begin{split}	
			& \underset{\bm W_k, t_k}{\mathop{\max}}\,\text {Tr}(\bm H_1^k\bm W_k)\gamma -\chi (	||\bm W_k||_*-||\bar {\bm W}_{k}(j)||)\\
			&\text{s.t.}~\bar C_{2-1},\bar C_{3-5}, \bar C_{4-4}, \bar C_6.	
		\end{split}
	\end{equation} 
Although problem (\ref{sp6}) is not jointly concave with respect to $\bm W_k$ and $t_k$, (\ref{sp6}) is a convex optimization problem which can be solved optimally by using some commercial solvers such as CVX for a given $t_k$. Then the optimal $t_k$ can be obtained by one-dimensional  exhaustive search over $t_k$. Specifically, the upper and lower bounds of $t_k$ can be chosen as \cite{LiZhangQin}, respectively, 
	\begin{equation} \label{19} 	
		\left\{ \begin{aligned}
			&t_k^{\text{ub}}=\max_{\bm W_k} \text{Tr}(\bm H_0 \bm W_k)=\lambda_k^{\max}(\bm H_0),\\ 
			& t_k^{\text{lb}}=\min_{\bm W_k} \text{Tr}(\bm H_0 \bm W_k)=\lambda_k^{\min}(\bm H_0).\\ 
		\end{aligned} \right.
	\end{equation}
The detail of this algorithm is shown in \textbf{Algorithm 2}.  Here, the computational complexity  for \textbf{Algorithm 2} is evaluated. Before solving problem (\ref{sp6}), it needs $\mathcal{O}(T)$ to determine the variables $t_k$.  Given the number of
constraints, interior point methods with a matrix variable $\bm W_k$ of size $M\times M$ will take $\mathcal O(\sqrt{M} \ln(1/\epsilon))$ iterations, with each iteration requiring $\mathcal{O}(M^6)$ arithmetic operations for the worst case. Letting the maximum iteration number for SCA  is $J$,  the total computation complexity for beamforming design is $\mathcal{O}(TJM^{6.5}\ln(1/\epsilon))$.

In what follows,  the remaining issue of tag selection for both problem (\ref{p1}) and problem (\ref{p2a}) is considered, which is a 0/1 nonlinear optimization problem. There have
been quite a few existing methods to solve such NP-hard problems, such as greedy algorithm, simulated annealing, penalty dual decomposition (PDD) method \cite{ShiHong}, and some compressive sensing methods like LASSO \cite{OsbornePresnell} and matching pursuit \cite{BeryhiVagollari}. In this paper, the greedy algorithm is applied. The detail of the greedy-based algorithm is shown in \textbf{Algorithm 3}.  Furthermore, when the corresponding beamforming designs are  determined via \textbf{Algorithm 1} and \textbf{Algorithm 2}, respectively, it costs $\mathcal{O}(K)$ for tag selection in \textbf{Algorithm 3}.

In summary,  the absolute CI can be achieved in this case, but a higher computation complexity  is incurred than the the case with the consensual CI.  In other words, there is a tradeoff between CI strictness and computation complexity.

\begin{table}[t]
	\centering
	\begin{tabular}{>{\raggedleft}p{0.5cm}p{7.5cm}}
		\toprule
		\multicolumn{2}{l}{\textbf{Algorithm 3} The Greedy Algorithm for Tag Selection}\\
		\toprule
		\multicolumn{2}{l} {\textbf{Input:} $K$, $M$, $N$, $\alpha_k$, $h^k_{\text{ST}}$, $\bm h_{\text{SR}}$, $\bm h^k_{\text{TR}}$, $\sigma_w^2$, $\sigma_s^2$, $ \xi^{\max}$, $\zeta^{\max}$.}\\
		\multicolumn{2}{l} {\textbf{Output:} $\beta_k$ and $\hat \beta_k$.}\\
		1: &\textbf{for} $k=1:K$ \\
		2: &~~~Solve problem (\ref{p1e}), obtain $\bm {v}_k^*$ via  \textbf{Algorithm 1}.\\
		3: &~~~Update $S_k=|(\bm {v}_k^*)^H\bm h_1|^2\gamma$.\\
		4: &~~~Solve problem (\ref{sp6}), obtain $\bm {\hat v}_k^*$ via  \textbf{Algorithm 2}.\\
		5: &~~~Update $Q_k=|(\bm {\hat v}_k^*)^H\bm h_1|^2\gamma$.\\
		6: &\textbf{end for}\\
        7:& Obtain the optimal tag selection to (\ref{p1e})  as $\beta_{k^*}=1$, where $k^*=\arg \underset{k}{\mathop{\max}}\ S_k$. \\
        8:& Obtain the optimal tag selection to (\ref{sp6})  as $\hat \beta_{k^*}=1$, where $k^*=\arg \underset{k}{\mathop{\max}}\ Q_k$. \\
		\toprule
	\end{tabular}
\end{table}

\subsection{Extension to a Multi-antenna PS }
	
	We have investigated the CI availability for a BackCom system with the single-antenna PS in the previous subsections. In what follows, we plan to explore the CI availability for a multiple-input and multiple-output (MIMO)-based BsckCom system by extending the PS with $Q$ transmit antennas. In contrast to the previous sections, due to the implementation of the multi-antenna PS, it is necessary to take into account the handling of the transmit beams. Building on this, the received signal of the reader with and without DL can be extended to
	\begin{equation}  \label{m1}
		\left\{ \begin{aligned}
			&y_k(n){=}\bm v_k^H \left( (\bm H_{\text{SR}})^H{+}\alpha_k B_k \bm h_{\text{TR}}^{k}(\bm {\bar h}_{\text{ST}}^{k})^H \right) \bm x_k s(n){+}{\bm v_k^H \bm w(n)},  \\		
			&\bar y_k(n){=}\bm v_k^H   \bm h_{\text{TR}}^{k}(\bm {\bar h}_{\text{ST}}^{k})^H \bm x_k \alpha_k B_k  s(n){+}{\bm v_k^H \bm w(n)},  \\	
		\end{aligned} \right.
	\end{equation}
	where $\bm {\bar h}_{\text{ST}}^{k} \in \mathbb{C}^{Q\times1}$,  $\bm H_{\text{SR}} \in \mathbb{C}^{Q\times M}$, and $ \bm x_k \in \mathbb{C}^{Q\times1}$ denotes the transmit beamforming vector, which satisfies $||\bm x_k||^2=\sigma_s^2$.

	Similar to the analysis in the previous subsection, by redefining $\delta_0^k$, $\bar \delta_0^k$, $\delta_1^k$, and $\bar \delta_1^k$ as  
	\begin{equation} \label{m2} 	
		\delta_i^k=	\left\{ \begin{aligned}
			&  |\bm v_k^H \bm g_0 \bm x_k|^2+\sigma_w^2||\bm v_k||^2,i=0, \\
			&  |\bm v_k^H \bm g_1^k \bm x_k|^2+\sigma_w^2||\bm v_k||^2, i=1, \\
		\end{aligned} \right.
	\end{equation}
	\begin{equation} \label{m3} 	
		\bar \delta_i^k=	\left\{ \begin{aligned}
			&  |\bm v_k^H \bm {\bar g}_0 \bm x_k|^2+\sigma_w^2||\bm v_k||^2,i=0,\\
			&  |\bm v_k^H \bm {\bar g}_1^k \bm x_k|^2+\sigma_w^2||\bm v_k||^2, i=1, \\
		\end{aligned} \right.
	\end{equation}
	the KLD and the DEP under different hypotheses with  the DL can still be obtained as (\ref{12}) and (\ref{12a}).  Correspondingly, the KLD and the DEP under different hypotheses without the DL can still be represented by (\ref{7}) and (\ref{7a}), where $\bm g_0=(\bm H_{\text{SR}})^H$, $\bm {\bar g}_0= \bm 0$, $\bm g_1^k=(\bm H_{\text{SR}})^H{+}\alpha_k \bm h_{\text{TR}}^{k}(\bm {\bar h}_{\text{ST}}^{k})^H$, and $\bm {\bar g}_1^k= \alpha_k \bm h_{\text{TR}}^{k}(\bm {\bar h}_{\text{ST}}^{k})^H$.

	\subsubsection{Extension for Consensual CI}
	Using the same transformations as (\ref{c1}) and (\ref{c2a}), problem (\ref{p1b}) can be extended to 
	\begin{equation} \label{pm1} 
		\begin{aligned}	
			& \underset{\bm v_k, \bm x_k}{\mathop{\max}}\,  \frac {|\bm v_k^H \bm {g}_1^k \bm x_k|^2}{\sigma_w^2}  \\ 
			&\text{s.t.}~C_{2}: ||\bm v_k||^2=1, \\
			&\quad ~~C_{3}:D_{\text{KL}}^k(p_0||p_1)\ge D^{\min},\\ 
			&\quad ~~C_{4}: D_{\text{KL}}^k(\bar p_0||\bar p_1)\ge E^{\min},\\ 
			&\quad ~~C_{5}: ||\bm x_k||^2=\sigma_s^2.\\
		\end{aligned}
	\end{equation}

	\subsubsection{Extension for Evolved CI}
	Referring to the transformation of (\ref{p2ba}) and (\ref{p2bb}), problem (\ref{p2b}) can be extended to 
	\begin{equation} \label{pm2} 
		\begin{aligned}	
			& \underset{\bm {\hat v}_k, \bm {\hat x}_k}{\mathop{\max}}\,  \frac {|\bm {\hat v}_k^H \bm {g}_1^k \bm {\hat x}_k|^2}{\sigma_w^2}  \\ 
			&\text{s.t.}~\bar C_2: ||\bm {\hat v}_k||^2=1, \\ 
			&\quad ~~\bar C_{3}: \Delta D_{\text{KL}}^k\ge 0,\\ 
			&\quad ~~\bar C_{4}:D_{\text{KL}}^k(\bar p_0||\bar p_1)\ge E^{\min},\\ 
			&\quad ~~\bar C_{5}: ||\bm {\hat x}_k||^2=\sigma_s^2.\\
		\end{aligned}
	\end{equation}

	\subsubsection{Algorithm Design}
	
	\begin{table}[t]
		\centering
	\begin{tabular}{>{\raggedleft}p{0.5cm}p{7.5cm}}
				\toprule
				\multicolumn{2}{l}{\textbf{Algorithm 4} The Alternating Iteration Algorithm}\\
				\toprule
				\multicolumn{2}{l} {\textbf{Input:} $K$, $M$, $N$, $Q$, $\alpha_k$, $\bm {\bar h}^k_{\text{ST}}$, $\bm H_{\text{SR}}$, $\bm h^k_{\text{TR}}$, $\sigma_w^2$, $\sigma_s^2$, $ \xi^{\max}$, }\\
				\multicolumn{2}{l} {\quad \quad \quad $\zeta^{\max}$.}\\
				\multicolumn{2}{l} {\textbf{Set:} Set the alternating iteration index $s=1$.}\\
				\multicolumn{2}{l} {\textbf{Initialize:} Set initial points $\bm v_k(1)$ and $\bm {\bar v}_k(1)$.}\\
				\multicolumn{2}{l} {\textbf{Output:} $\bm x_k$, $\bm v_k$, $\bm {\hat x}_k$ and $\bm {\bar v}_k$.}\\
				&\textbf{For consensual CI via solving problem (\ref{pm1})}\\
				1: &\textbf{repeat}  \\
				2: &~~~With given $\bm v_k(s)$, obtain $\bm x_k (s+1) $ by solving\\
				&~~~ problem (\ref{pm1}) via  \textbf{Algorithm 1}.\\
				3: &~~~With given $\bm x_k(s+1)$, obtain $\bm v_k (s+1) $ by solving\\
				&~~~ problem (\ref{pm1}) via  \textbf{Algorithm 1}.\\
				4:&~~~Update $s=s+1$.\\
				5: &\textbf{until convergence}  \\
				&\textbf{For evolved CI via solving problem (\ref{pm2})}\\
				1: &\textbf{repeat}  \\
				2: &~~~With given $\bm {\hat v}_k(s)$, obtain $\bm {\hat x}_k (s+1) $ by solving\\
				&~~~ problem (\ref{pm2}) via  \textbf{Algorithm 2}.\\
				3: &~~~With given $\bm {\hat x}_k(s+1)$, obtain $\bm {\hat v}_k (s+1) $ by solving\\
				&~~~ problem (\ref{pm2}) via  \textbf{Algorithm 2}.\\
				4:&~~~Update $s=s+1$.\\
				5: &\textbf{until convergence}  \\
				\toprule
		\end{tabular}
	\end{table}

	Compared with problem (\ref{p1b}) and problem (\ref{p2b}) in the previous subsections,  problem (\ref{pm1}) and problem (\ref{pm2}) seem to be more thorny beyond the direct application of the proposed algorithms. A turnaround in this deadlock is possible thanks to the alternating optimization method. Taking problem  (\ref{pm1})  as an example, when $\bm x_k$ is fixed, the resulting problem can be regarded as formally equivalent to that of problem (\ref{p1b}), which can be solved using  \textbf{Algorithm 1}. Likewise, when $\bm v_k$ is fixed, the resulting problem can be solved by the same way. Then, using the alternating iteration method, problem  (\ref{pm1}) is solved when the results of the alternation converge. Not surprisingly, problem (\ref{pm2}) can also be solved using alternating optimization. Similarly, the desired solution for problem (\ref{pm2})  can be obtained by alternately using  \textbf{Algorithm 2} by fixing $\bm {\hat x}_k$ and $\bm {\hat v}_k$ separately. The detail of the proposed alternating iteration algorithm is shown in \textbf{Algorithm 4}.
	
	Note that \textbf{Algorithm 4} applies under the presumption that the $k$-th tag has already been selected. Then, how to select the optimal tag for transmission can be done according to the proposed \textbf{Algorithm 3}.
	
	\subsection{ Feasible Region for Input SNR and CI Angle}
	
In previous subsections, we have investigated and analyzed the CI availability for the multi-antenna BackCom system with the beamforming design. However, the ointment is that some analytical insights have not been obtained due to its intractability. In this subsection, in order to gain more insights into the CI, we consider degrading the system with SIMO/MIMO into a one with single input single output (SISO) to facilitate our analysis for the CI. In particular, according to (\ref{0}) and (\ref{11a}), the received signal of the reader with and without the DL can be degraded  into
	\begin{equation}  \label{s1}
			\left\{ \begin{aligned}
					&y_k(n)= h_{\text{SR}} s(n)+B_kh^k_{\text{STR}} s(n)+w(n),\\
					&\bar y_k(n)=B_kh^k_{\text{STR}} s(n)+w(n),
			\end{aligned} \right.
	\end{equation}
where $h_{\text{SR}}$ and $h_{\text{STR}}^k=\alpha_kh_{\text{ST}}^kh_{\text{TR}}^k$ denote the channel coefficients of the DL and the BL, respectively. $h_{\text{ST}}^k$ and $h_{\text{TR}}^k$ are the channel coefficients of the  forward link and backward link for the $k$-th tag. $w(n)$ is the noise at the reader with zero mean and  variance $\sigma_w^2$.

Similar to the previous analysis on the KLDs under two hypotheses (i.e., $\mathcal H_0$ and $\mathcal H_1$),  to make the DLI  constructive, the following KLD constraint for the $k$-th tag need to be satisfied 
	\begin{equation} \label{s3}
		\begin{aligned}
			&\Delta D_{\text{KL}}^k=N\left[\left(\ln  \frac{\delta_1^k}{\delta_0}+\frac{\delta_0}{\delta_1^k}\right)-\left( \ln \frac{\bar \delta_1^k}{\bar \delta_0}+\frac{\bar \delta_0}{\bar \delta_1^k}\right)  \right]\ge 0.
		\end{aligned}
	\end{equation}
	where $\delta_0=\sigma_s^2|h_0|^2+\sigma_w^2$, $\bar \delta_0=\sigma_s^2|\bar h_0|^2+\sigma_w^2$, $\delta_1^k=\sigma_s^2|h_1^k|^2+\sigma_w^2$,  $\bar \delta_1^k=\sigma_s^2|\bar h_1^k|^2+\sigma_w^2$, $h_0=h_{\text{SR}}$, $\bar h_0=0$, $h_1^k=h_{\text{SR}}+h^k_{\text{STR}}$, and $\bar h_1^k=h^k_{\text{STR}}$.

On the other hand, by degrading $C_{4-2}$, we have 
	\begin{equation} \label{s4} 	
		 \begin{aligned}
N\left( \ln \frac{\bar \delta_1^k}{\bar \delta_0}+\frac{\bar \delta_0}{\bar \delta_1^k}-1\right)\ge E^{\min}_{k}.
		\end{aligned} 
	\end{equation}

To reveal the CI region,  theorem 3 is proposed.

\textbf{\textit{Theorem 3:}} The evolved CI can be achieved when the input SNR ($\gamma$) is satisfied, i.e.,
		\begin{equation}
		\begin{aligned} \label{s3c}
			& \frac{F(g^{\min})-1}{|h^k_{\text{STR}}|^2}\le \gamma \le \frac{(h_{\text{SR}})^Hh^k_{\text{STR}}+h_{\text{SR}}	(h^k_{\text{STR}})^H}{|h_{\text{SR}}|^2|h^k_{\text{STR}}|^2}.\\	
		\end{aligned}
	\end{equation}

\textit{\textbf{Proof:}} Please see Appendix C.
	
	\textbf{\textit{Remark 2:}} From (\ref{s3c}), it can be seen that the evolved CI region for $\gamma$ is obtained. To be specific, the introduced DLI is fully constructive for BackCom systems when $\gamma$ lies within this region. Besides, as $g^{\min}$ becomes larger, the lower bound for $\gamma$ will become more strict accordingly, and the system may not be able to achieve the desired performance with a small $\gamma$. On the other hand, when the channel gain of the BL becomes small, it will also increase the lower bound of $\gamma$, which shows that more transmit power is needed to construct the CI region with a lower BL gain. However, once the required $\gamma$ exceeds its upper bound, the evolved CI will cease to exist.

To further show the relationship between the BL and the DL, we give theorem 4.

\textit{\textbf{Theorem 4:}}  Letting $\theta_k\triangleq \angle  \left\langle h_{\text{SR}}, h^k_{\text{STR}}\right\rangle$ denote the angle between the DL and the BL, the resulting DLI is constructive when $\theta_k$ satisfies the following condition, i.e.,
	\begin{equation} \label{s3d}
		\begin{aligned}
		 0\le \theta_k &\le \min\left\{\frac{\pi}{2},\arccos \frac{1}{2}\gamma{|h_{\text{SR}}||h^k_{\text{STR}}|}\right\},\\
		\end{aligned}
	\end{equation}
where $\gamma\ge \frac{F(g^{\min})-1}{|h^k_{\text{STR}}|^2}$.
	
	\textbf{\textit{Proof:}} Please see Appendix D.
	
\textbf{\textit{Remark 3:}} It can be seen that  the CI angle can be expressed as (\ref{s3d}). In particular, $\arccos \frac{1}{2}\gamma{|h_{\text{SR}}||h^k_{\text{STR}}|}$  is gradually narrowing the gap with $\frac{\pi}{2}$ with the decreasing $\gamma$, $|h^k_{\text{STR}}|$, and $|h_{\text{SR}}|$. When $\gamma^{\min}= \frac{F(g^{\min})-1}{|h^k_{\text{STR}}|^2}$ holds, the maximum CI angle can be achieved by substituting $\gamma^*$ into (\ref{s3d}), i.e., $\theta_k^{\max}=\arccos \frac{|h_{\text{SR}}|}{2}\frac{F(g^{\min})-1}{|h^k_{\text{STR}}|}$. It is noted that $\theta_k^{\max}$ is decreased with the increasing $g^{\min}$ and the increasing $|h_{\text{STR}}^k|$. That is to say, the maximum CI angle  mainly depends on the minimum KLD requirement with the determined channel gains. The simulation results below also verify what we have analysed.

	\section{Simulation Results}
	
	In this section, we present the simulation results of our proposed algorithms for BackCom networks. We assume that there are  one PS, one reader with 4 antennas (i.e., $M=4$), and 5 tags (i.e., $K=5$)  in the proposed system. The distances between the PS and tags,  the PS and the reader, as well as  the reader to tags are both within 5 meters.  $w_k$ is set to follow $\mathcal{CN} (0,0.03)$ \cite{TaoLiLiang2019}. Other parameters include $\alpha_k=0.8$, $\xi^{\max}=\zeta^{\max}=0.5$, and $T=J=L=10^3$.  The simulation results are evaluated based on the CVX package \cite{GrantBoyd}. 
	
	 Besides, we consider the distance-dependent path loss as large scale fading, and Rician fading as small scale fading for all channels \cite{XuGuLi2021,RamazaniAj}. For example, the channel between the PS and the reader is given by  $\bm h_{\text{SR}}=\sqrt{\frac{\kappa}{\kappa+1}} \bm h_{\text{SR}}^{\text{LoS}}+\sqrt{\frac{1}{\kappa+1}}\bm h_{\text{SR}}^{\text{NLoS}}$, where $\kappa=2.8$ is the Rician factor, $\bm h_{\text{SR}}^{\text{LoS}} \in \mathbb{C}^{M\times1}$ and $\bm h_{\text{SR}}^{\text{NLoS}} \in \mathbb{C}^{M\times1}$ are the line-of-sight (LoS) and non-line-of-sight (NLoS) components of  $\bm h_{\text{SR}}$, respectively.  Specifically, $\bm h_{\text{SR}}^{\text{LoS}}=[1, e^{-j\pi\sin(\theta_i)},\dots,e^{-j\pi(M-1)\sin(\theta_i)}]$, where $\theta_i$ is the the direction of the PS to the reader. $\bm h_{\text{SR}}^{\text{NLoS}}$ follows the standard Rayleigh fading.  The average power of $\bm h_{\text{SR}}$ is then normalized by $d_{\text{SR}}^{-\rho}$ , where $d_{\text{SR}}$ is the distance between the PS and the reader, and $\rho= 3$ is the path-loss exponent.
	
 In what follows, we will demonstrate the superiority of the proposed schemes from the perspectives of system performance and CI existence, respectively.
	
		\begin{figure*}[t]
		\centering
		\subfigure[The received SINR/SNR versus the transmit power of the PS.]
		{
			\includegraphics[width=3.2in]{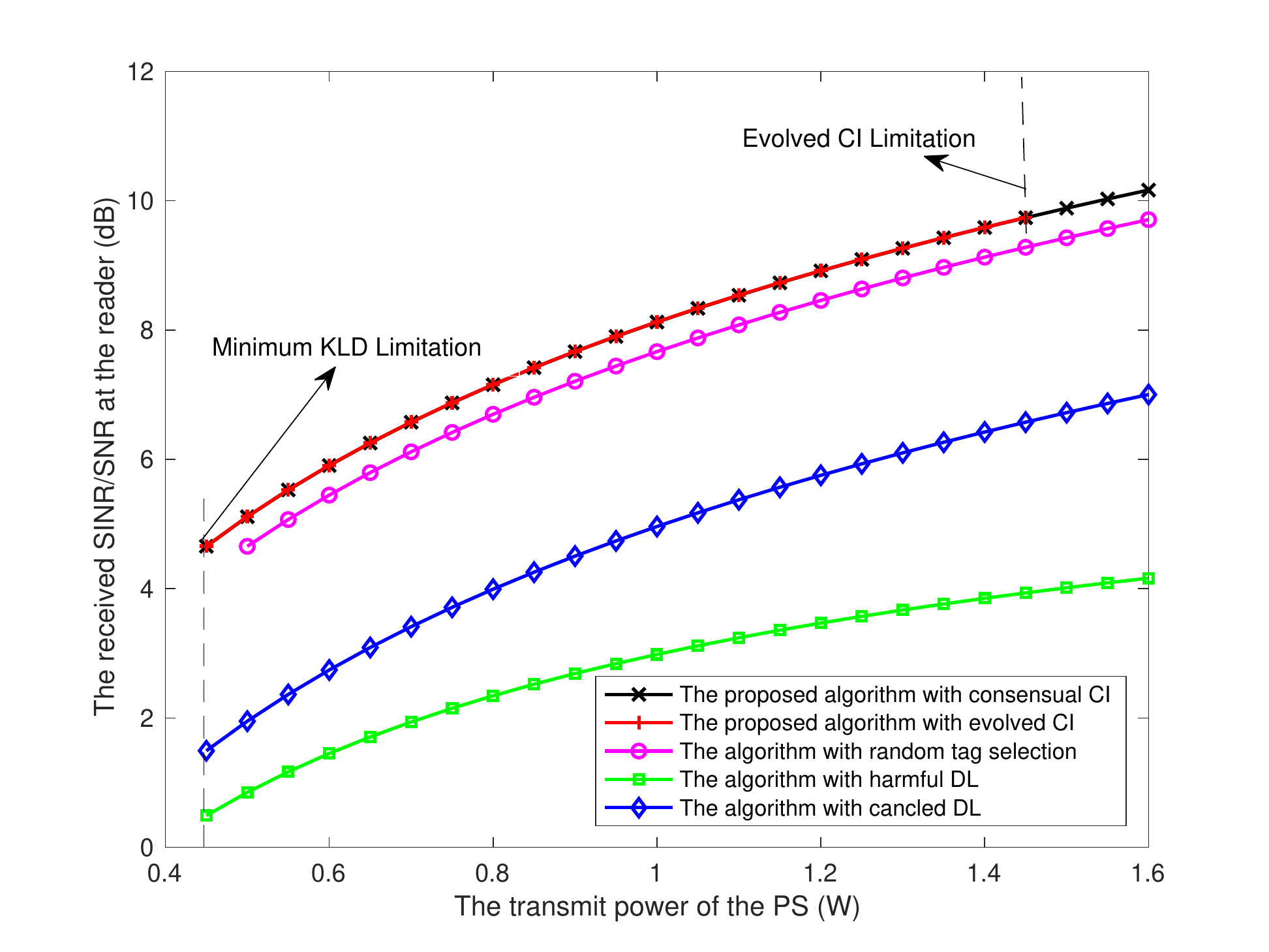} 
		}
		\subfigure[The KLD and the DEP versus the transmit power of the PS.]
		{
			\includegraphics[width=3.2in]{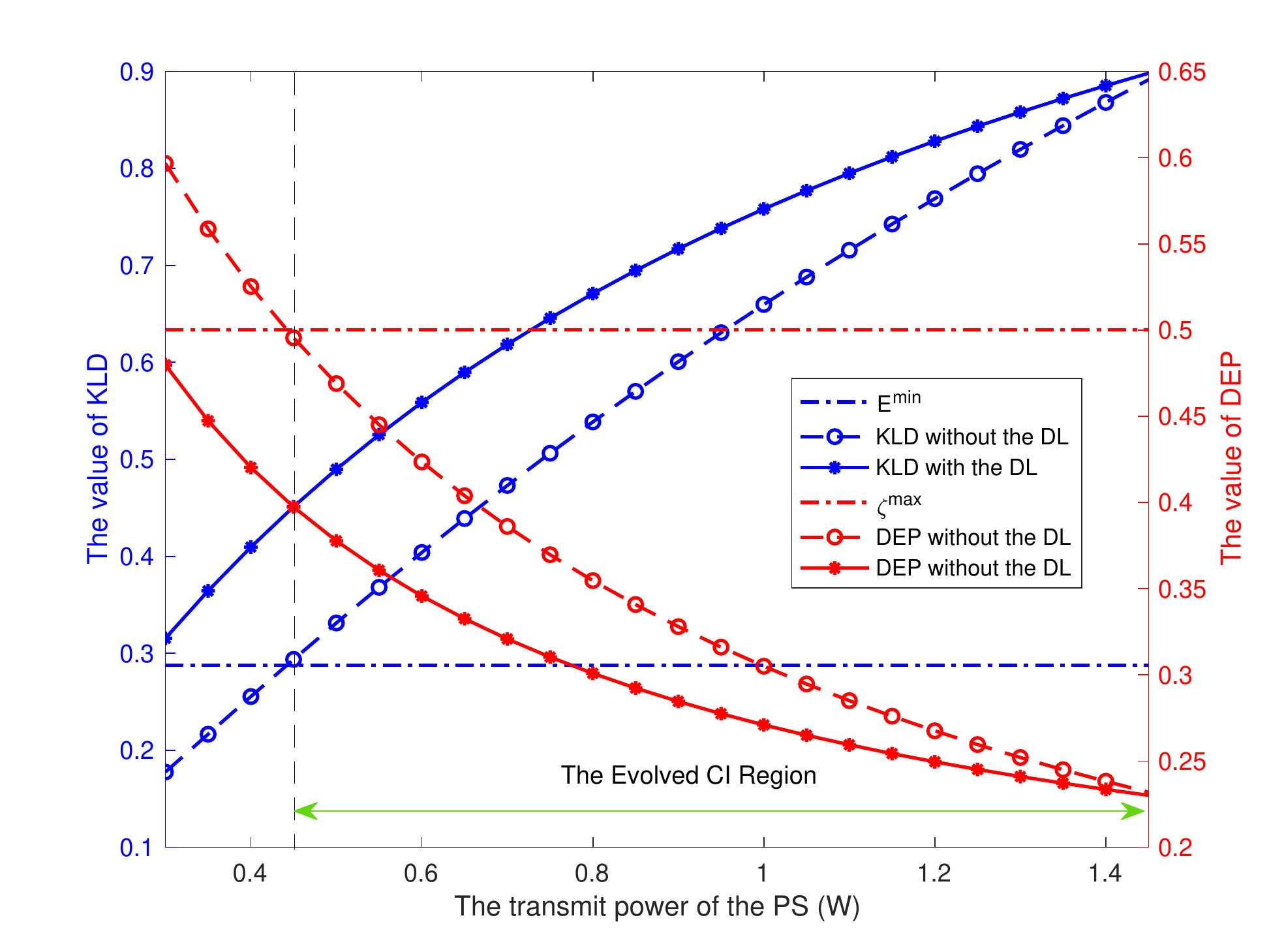} 
		}
		\DeclareGraphicsExtensions.
		\caption{The algorithm comparison and the evolved CI region analysis under the transmit power of the PS.}
		\label{fig3}
	\end{figure*}

\subsection{System Performance with SIMO and MIMO}

In this subsection, we begin by investigating the system performance for SIMO- and MIMO-based BackComs, where the CI is guaranteed. For algorithm comparison, we define three benchmark algorithms, such as

\begin{itemize}

	\item \textit{\textbf{The algorithm with harmful DLI}}
	
	The DL, as usual, is regarded as the harmful interference in this algorithm, which means that the DL no longer assist in enhancing the backscattering signal. This also results in a new optimization problem and a new solution, which are detailed in  Appendix E.

	\item \textit{\textbf{The algorithm with canceled DLI}}
	
	This scheme is consistent with the existing assumption that the DLI can be perfectly canceled. That is to say, each tag can enjoy interference-free transmission. This scheme can be obtained from the proposed scheme by removing the CI constraint and the DL.

	\item \textit{\textbf{The algorithm with Random Tag Selection}}
	
     Different from the proposed algorithm, each tag is selected  randomly under this scheme. This scheme can be obtained from the proposed scheme by replacing the greedy algorithm with a random approach, which can reduce the computation complexity.
\end{itemize}

\begin{figure*}[h]
	\centering
	\subfigure[The received SINR/SNR versus the channel path-loss exponent.]
	{
		\includegraphics[width=3.2in]{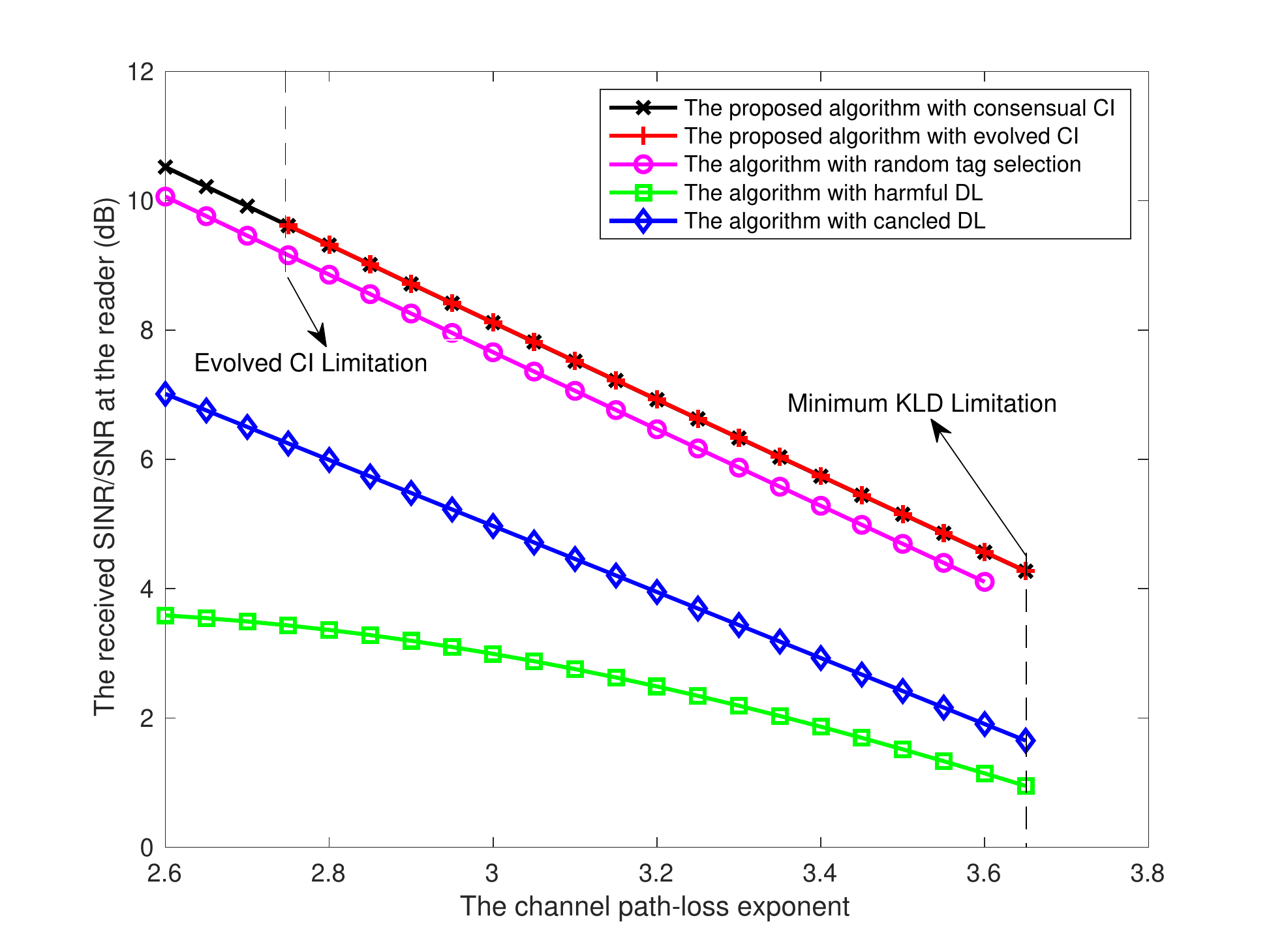} 
	}
	\subfigure[The KLD and DEP versus the channel path-loss exponent.]
	{
		\includegraphics[width=3.2in]{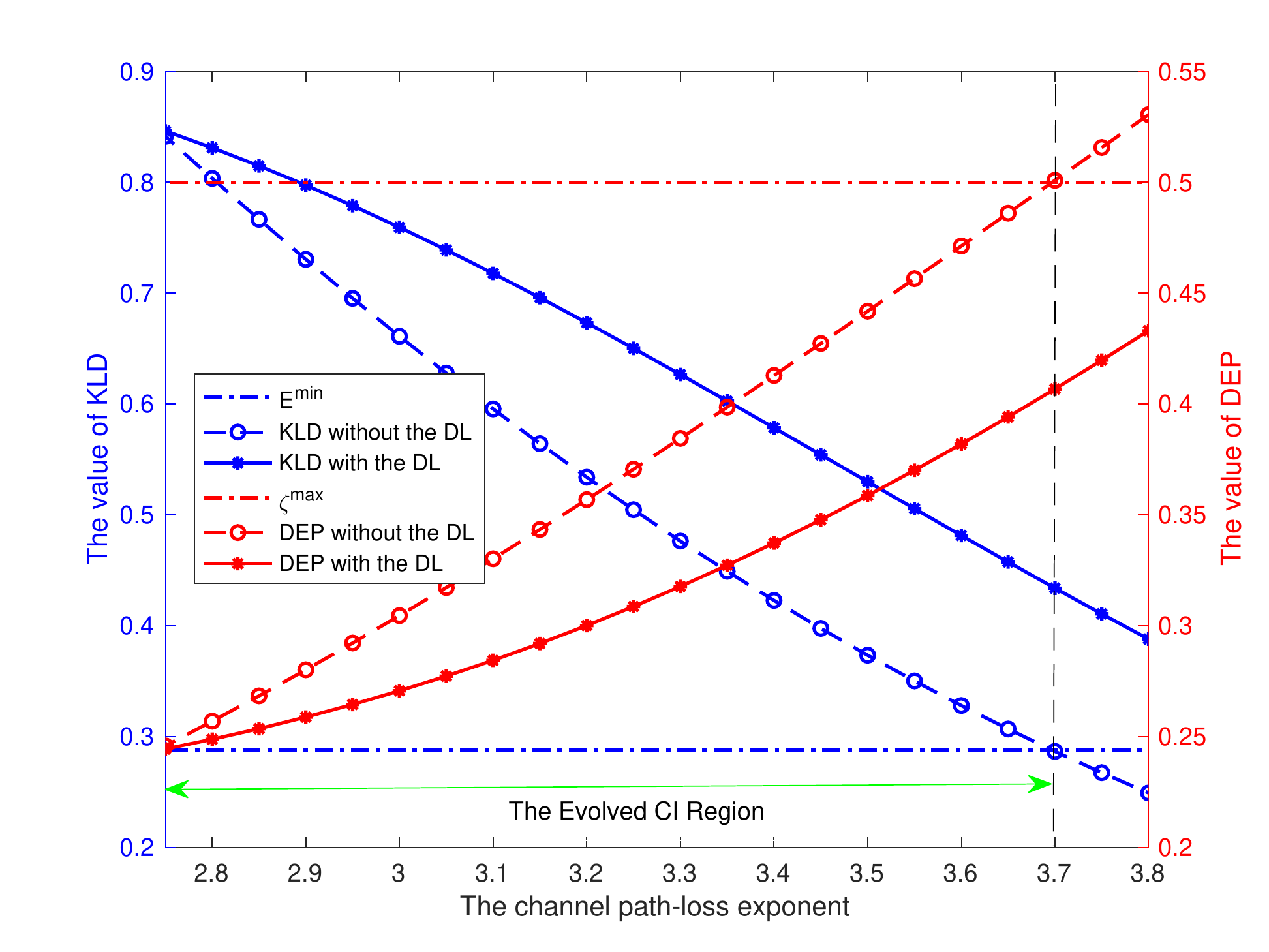} 
	}
	\DeclareGraphicsExtensions.
	\caption{The algorithm comparison and the evolved CI region versus the channel path-loss exponent.}
	\label{fig4}
\end{figure*}

The purpose is to evaluate the performance and effectiveness of the proposed algorithms under different aspects by comparing them with benchmark algorithms. On the one hand,  by comparing  the algorithm with harmful DL, we can measure the benefit of considering CI on the system performance. On the other hand, by comparing the  algorithm without the DL, we can evaluate the effectiveness of the proposed algorithms in enhancing the backscattered signal power. Besides, by comparing the  algorithm with random tag selection, we can evaluate the gap between optimal and random selection. In the following, we compare the proposed algorithms with these benchmark algorithms with respect to the transmit power of the PS, the number of receive antennas, and the channel gains, respectively, to demonstrate the performance and implications of the proposed algorithms.

The algorithm comparison and CI region analysis under the transmit power of the PS ($\sigma_s^2$) are shown in Fig. \ref{fig3}(a) and   Fig. \ref{fig3}(b), respectively. From Fig. \ref{fig3}(a), it can be seen that, unsurprisingly, the received SINR/SNR of all algorithms increases with the increase in $\sigma_s^2$. Noteworthy,  the received SINR/SNR of both the algorithms with the canceled DL and the harmful DL is significantly smaller than that of the proposed algorithms. This behavior is due to the fact that they lose the boost from the DL to enhance the strength of the backscattered signal, thus resulting in the SINR/SNR of them is unable to be raised. However, it has to be admitted that the performance of the proposed algorithm with evolved CI is as good as that with consensual CI , but its feasible domain is restricted in a region. For the lower bound of the feasible region, it is limited by the minimum KLD threshold (or the required DEP) for all algorithms, especially, a tag with the worse BL condition needs a larger transmit power to satisfy this requirement. And  For the upper bound of the feasible region, there is no limitation for other algorithms, except the proposed algorithm with evolved CI. The reason is that its upper bound is limited by the maximum evolved CI limitation (or the gap of KLDs). For a more intuitive interpretation of the CI region, Fig. \ref{fig3}(b) shows the KLD and the DEP versus $\sigma_s^2$. From the perspective of DEP, the DEP without the DL only meets the preset when the transmit power reaches a certain level. On the other hand, when the transmit power is too large, the DEP with the DL will exceed the DEP without the DL, which means that the DL is not fully constructive for BackComs. Although the changes of the KLD versus the transmit power show the opposite trend to the DEP, the construction of CI regions is the same end.

Fig. \ref{fig4} shows the algorithm comparison and CI region analysis under the channel path-loss exponent. Specifically, from Fig. \ref{fig4}(a),  we can see that the received SINR/SNR of all algorithms decreases with the decreasing channel path-loss exponent. This is because the corresponding channel gains are degraded with the decrease in channel path-loss exponent. Similarly, the performance of the proposed algorithms is better than that of the other algorithms. The reason is that the DL is able to enhance the BL so that the backscattering capacity can be improved. Besides, the feasible region of the proposed algorithm with evolved CI is limited in order to meet the requirements of the evolved CI. As a supplementary for the evolved CI region, Fig. \ref{fig4}(b) shows the formation of the CI region from the perspective of the KLD and the DEP. It can be seen that the KLD with the DL is smaller than that without the DL when the channel path-loss exponent is less than 2.75, which means that the DL is harmful to the BackCom system. On the other hand, the KLD without the DL can not achieve the minimum threshold when the channel path-loss exponent is larger than 3.7, which means that the essential detection requirement can not be satisfied. The same analysis can apply to the DEP.

\begin{figure}[t]
	\centerline{\includegraphics[width=3.2in]{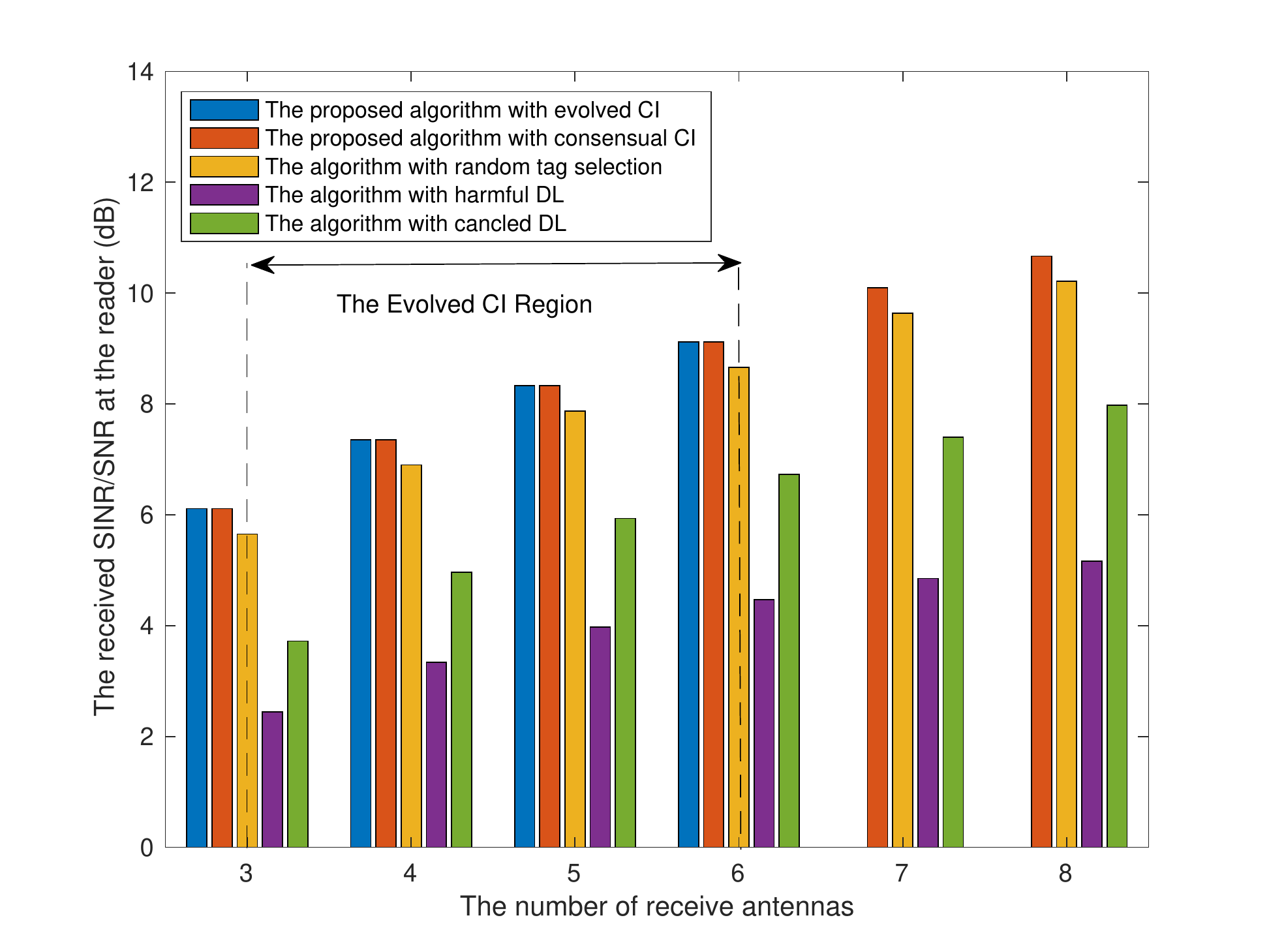}}
	\caption{The received SINR/SNR versus the number of receive antennas.}
	\label{fig5}
\end{figure}

Fig. \ref{fig5} shows the received SINR/SNR versus the number of receive antennas ($M$). It is seen that as $M$ increases, the received SINR/SNR of all algorithms increases gradually, which is attributed to the improvement in beamforming gain due to the increase in $M$. Although the algorithm with random selection has the advantage of low complexity compared to the proposed algorithm, the system performance can not be guaranteed. Moreover, the received SINR/SNR of the proposed algorithms is higher than that of both the algorithms with the canceled DL and the harmful DL since the DL can be borrowed to reinforce the backscattering transmission in this region. It is noted that the evolved CI region under the proposed algorithm can not be built when $M$ is too small and too large. This is because when $M$ is small, the DEP requirement for the proposed algorithm cannot be satisfied. On the other hand, when $M$ is relatively large, the signal from the DL is too strong to keep the DLI constructive, which can be explained through the same reason as Fig. \ref{fig3} and Fig. \ref{fig4}.

	\begin{figure}[t]
	\centerline{\includegraphics[width=3.2in]{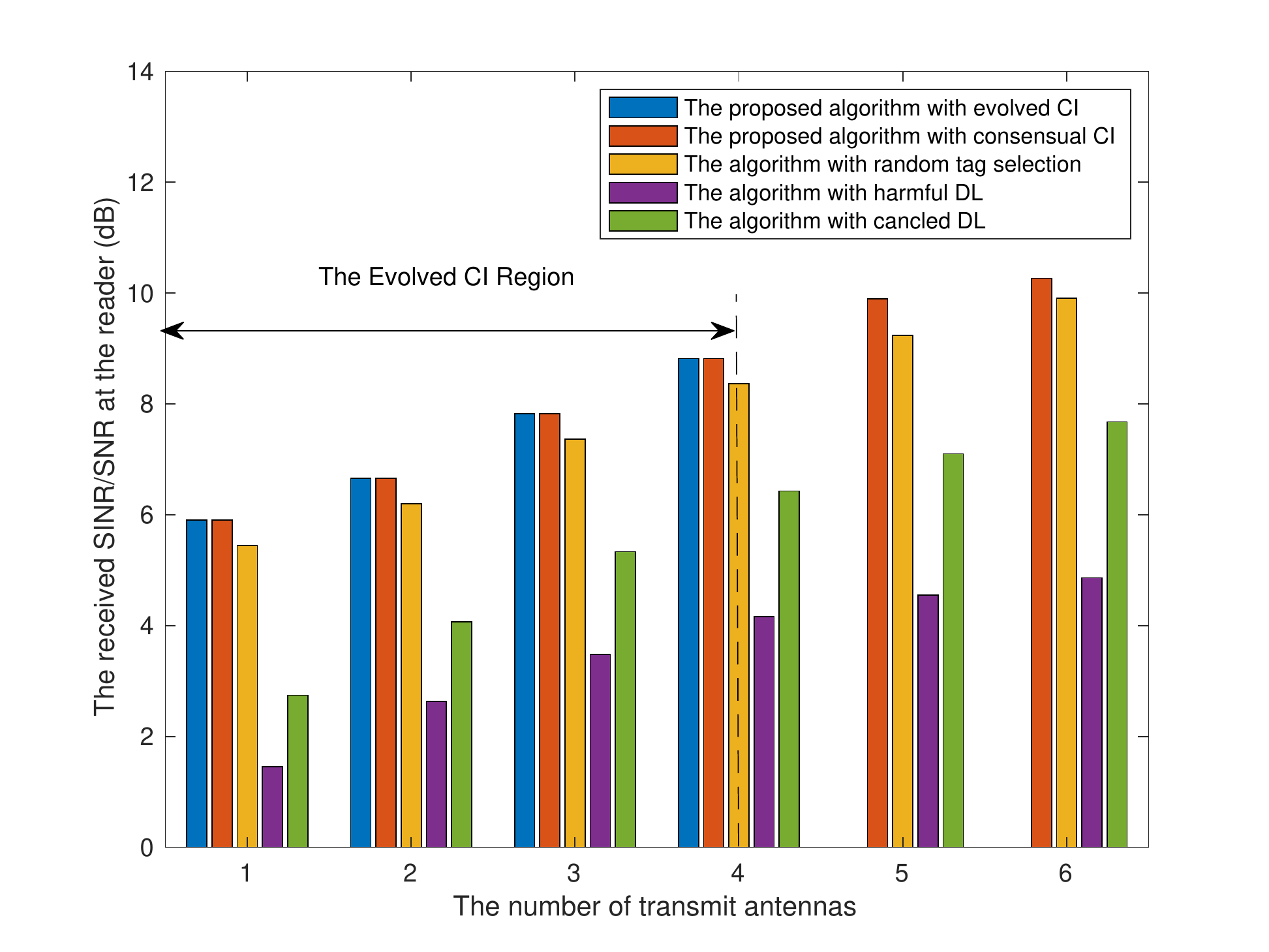}}
	\caption{The received SINR/SNR versus the number of transmit antennas.}
	\label{fig6}
\end{figure} 
	
	By extending the single-antenna PS to a multi-antenna one, as shown in Fig. \ref{fig6}, we numerically evaluate the system performance with MIMO. Specifically,  Fig. \ref{fig6} shows the received SINR/SNR versus the number of transmit antennas ($Q$) with $\sigma_s^2=0.6$ W. As $Q$ increases, the received SINR/SNR of all algorithms increases gradually, which can be attributed to an increase in beamforming gain.  Besides, the received SINR/SNR of the proposed algorithms is higher than that of both the algorithms with canceled DL and the harmful DL, since the DL can enhance the BL. It is noted that the evolved CI region under the proposed algorithm is beingless when $M$ is large than 4. It is due to the strong signal from the DL that it is impossible to keep the DLI constructive when $Q$ is relatively large for the same reason outlined in Fig. \ref{fig5}.  

\begin{figure}[t]
	\centering
	\subfigure[The transmit power versus the acceptable DEP  ($\zeta^{\max}$).]
	{
		\includegraphics[width=3.2in]{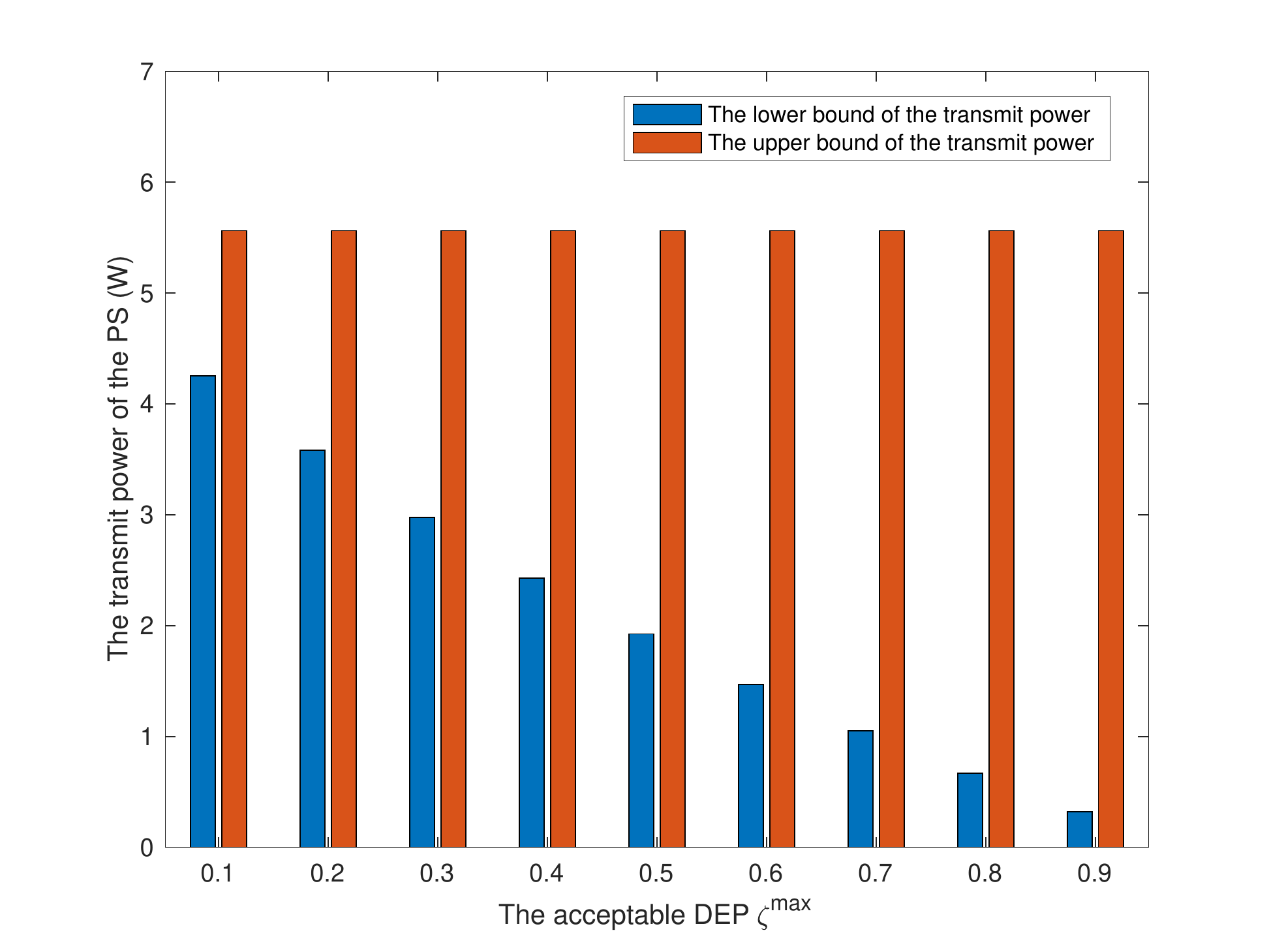} 
	}
	\subfigure[The CI angle versus the acceptable DEP  ($\zeta^{\max}$).]
	{
		\includegraphics[width=3.2in]{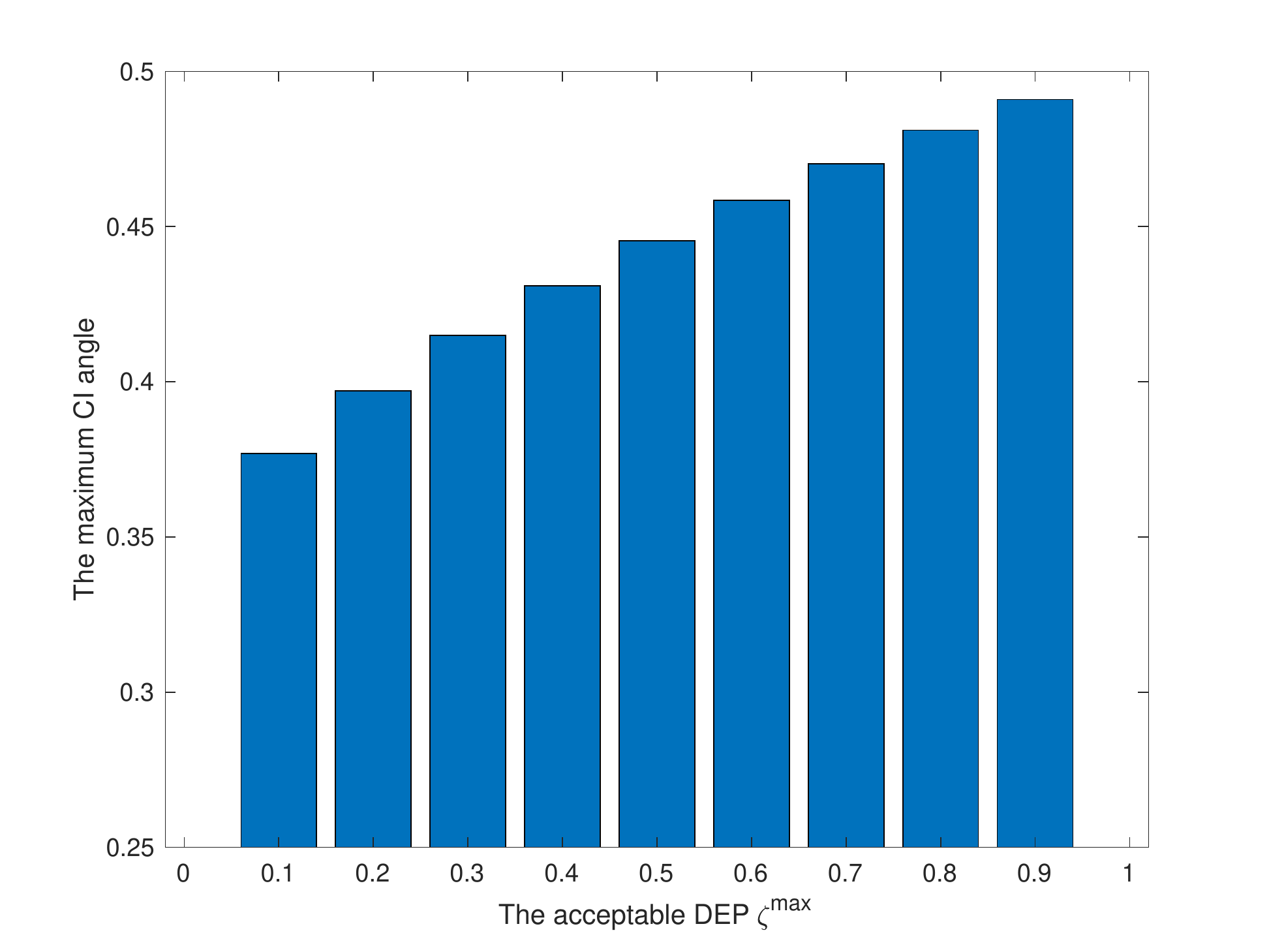} 
	}
	\DeclareGraphicsExtensions.
	\caption{The transmit power and the CI angle versus the acceptable DEP.}
	\label{fig7}
\end{figure}

\begin{figure}[t]
	\centering
	\subfigure[The transmit power versus the channel path-loss exponent.]
	{
		\includegraphics[width=3.2in]{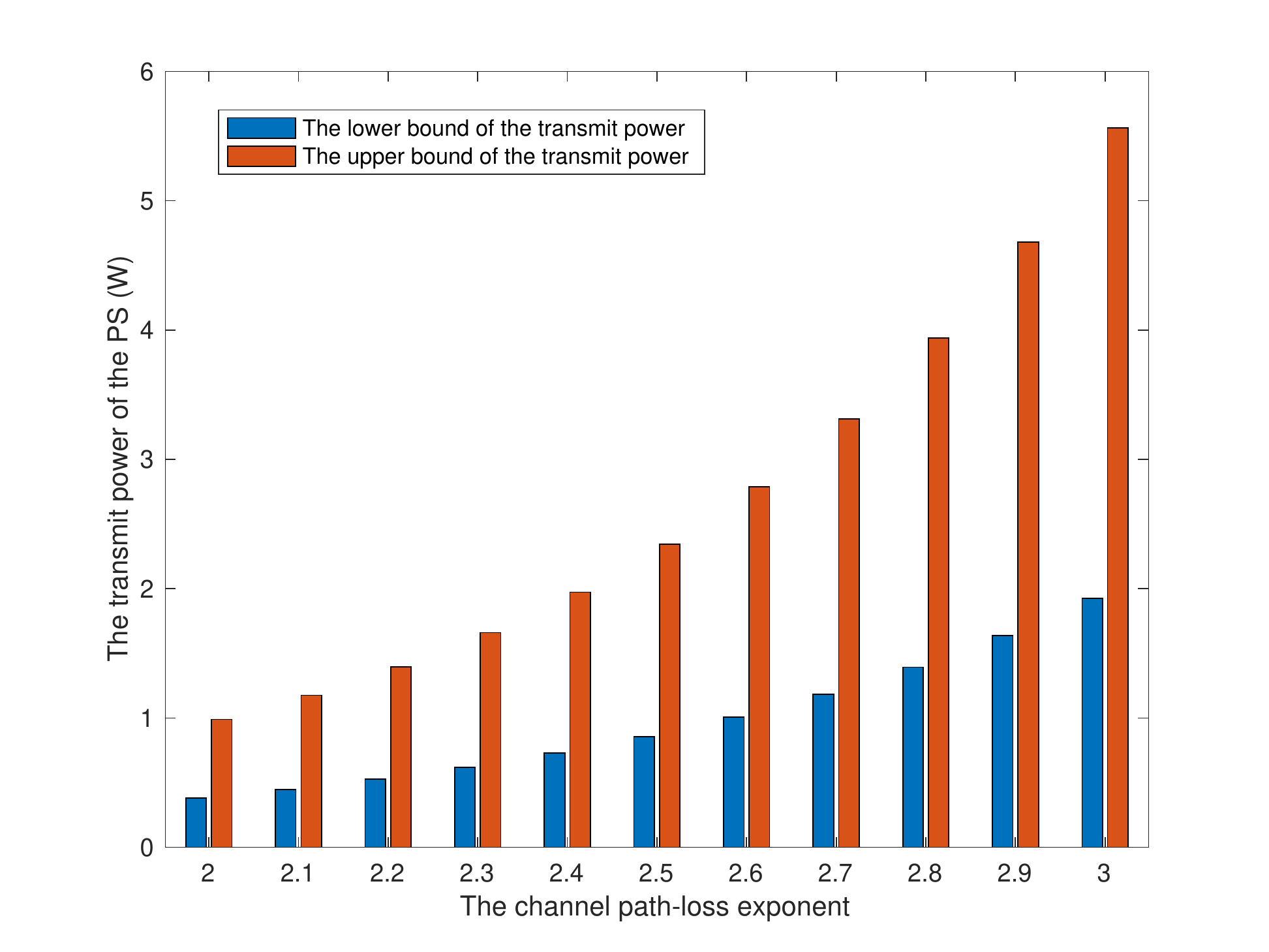} 
	}
	\subfigure[The CI angle versus the channel path-loss exponent.]
	{
		\includegraphics[width=3.2in]{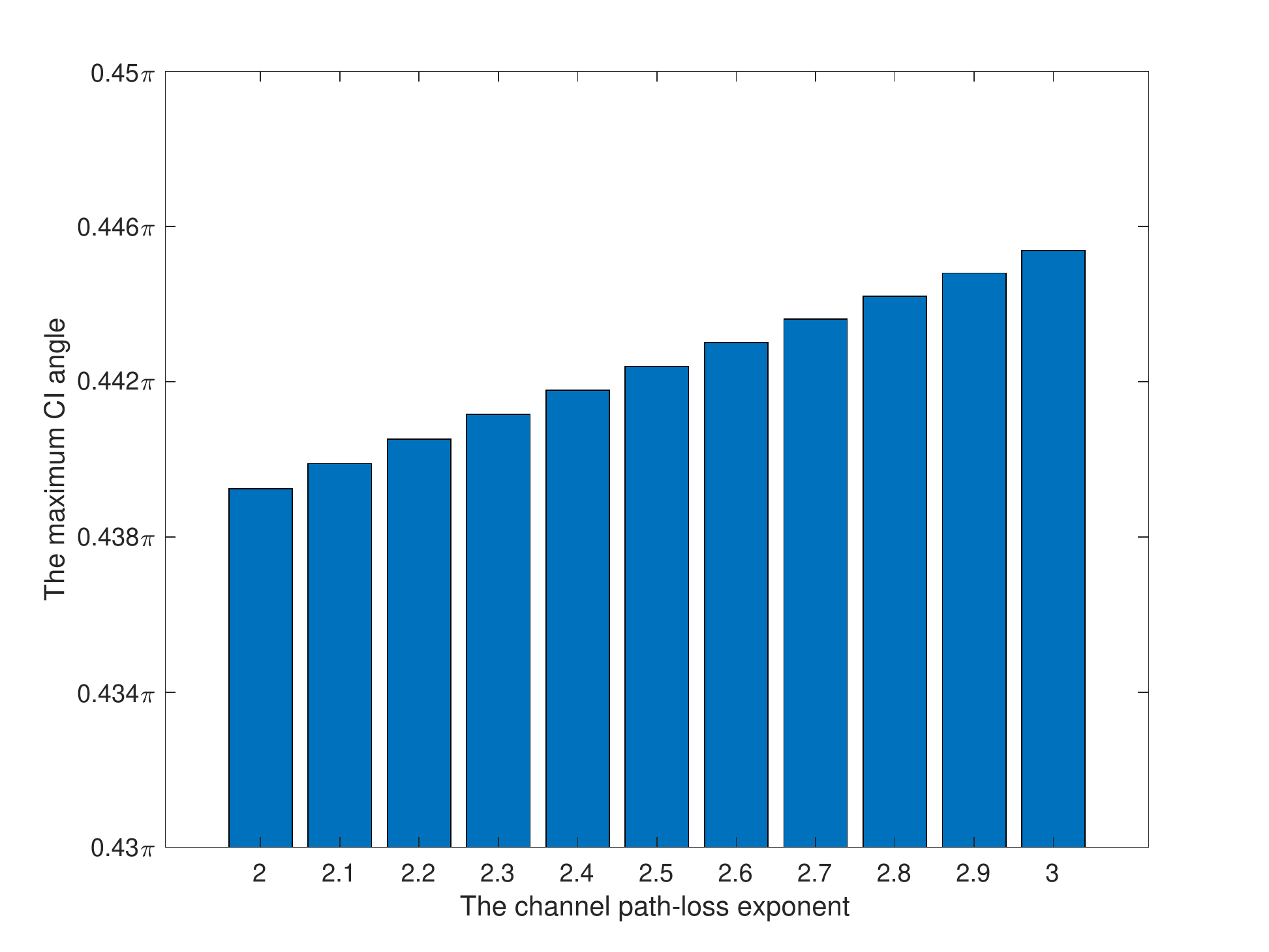} 
	}
	\DeclareGraphicsExtensions.
	\caption{The transmit power and the CI angle versus the channel path-loss exponent.}
	\label{fig8}
\end{figure}

\subsection{CI Existence with SISO}

In this subsection, we numerically evaluate the feasible region of the transmit power of the PS and the CI angle with $N=10$ for the evolved CI requirement, as shown in Fig. \ref{fig7} and \ref{fig8}, in SISO-based BackComs.

Fig. \ref{fig7} shows the transmit power and the CI angle versus the acceptable DEP ($\zeta^{\max}$). Fig. \ref{fig7}(a) shows the lower and upper bounds of the transmit power for building the CI under different $\zeta^{\max}$. As $\zeta^{\max}$ increases, only the lower bound of the transmit power decreases, while its upper bound is unaffected. The reason is that the increased $\zeta^{\max}$ makes $D_{{\min}}^{k}$ lower so that a small power is enough to satisfy the required KLD. This also illustrates that the CI region of transmit power becomes wider with a higher  $\zeta^{\max}$. Besides, it can be seen that, from Fig. \ref{fig7}(b),  the maximum CI angle decreases with the decreasing $\zeta^{\max}$. This means that when the DEP requirement becomes stricter, the feasible region of the CI angle is shrunk. In combination with Fig. \ref{fig6}(a), it shows a trade-off between the size of the CI region and the DEP, especially for the lower bound of the CI region.

The transmit power and the CI angle under the channel path-loss exponent are shown in Fig. \ref{fig8}(a) and Fig. \ref{fig8}(b), respectively. From Fig. \ref{fig8}(a), both the upper and lower bounds of the transmit power increase with the channel path-loss exponent, and the gap between them becomes more significant (i.e., the CI region of the transmit power becomes larger). 
The reason is that when the channel path-loss exponent becomes larger, the channel gain of both the DL and the BL decreases, which not only leads to a larger transmit power needed to satisfy the minimum KLD requirement but also enlarges the upper bound of the transmit power according to (\ref{s3c}). There is a weakly increasing trend of the maximum CI angle with the growth of the channel path-loss exponent as shown in Fig. \ref{fig8}(b). Since according to the expression of $\theta_k^{\max}$, $|h_{\text{SR}}|$ and $|h^k_{\text{STR}}|$ are in conflict, which attenuates the effect on maximum CI angle. In short, when the channel path-loss exponent is larger, the construction of the CI region has higher freedom.

\section{Conclusions}

This paper investigates and analyzes the CI region for BackComs with multiple tags and a multi-antennas reader, where tag selection and receive beamforming are taken into account. Specifically, a KLD-oriented approach is proposed to build the CI region for the DLI, and two optimization problems under different CI requirements with the single- and multi-antenna PSs, respectively, are formulated. Then, we reformulate these original problems with some transformation, and then propose the corresponding algorithms, respectively, which involve  two SCA-based alternative algorithms for beamforming design, and a greedy algorithm for tag selection. Furthermore, a closed-form expression for the CI angle is derived in the BackCom system with a single-antenna reader. Simulation results show that the performance of the proposed algorithm outperforms benchmark algorithms in terms of the received SNR, and confirm the derived channel angle for the CI.

\begin{appendices}

	\section{Proof of  Theorem 1}	

   Defining $f(x)=\ln x+\frac{1}{x}$, where $x>0$, the first-order derivative of $f(x)$ is  
   \begin{equation} \label{14} 	
   	\frac{	\partial f(x)}{\partial x}{=}\frac{1}{x}-\frac{1}{x^2}=\left\{ \begin{aligned}
   		& <0, 0<x<1, \\ 
   		& =0, x=1,\\ 
   		&>0, x>1.\\
   	\end{aligned} \right.
   \end{equation}
   
   Letting $x_k=\frac{\delta_1^k}{\delta_0^k}$ and $\bar x_k=\frac{\bar \delta_1^k}{\bar \delta_0^k}$, we have
   \begin{equation} \label{15} 	
   	\left\{ \begin{aligned}
   		&x_k{=} \frac{|\bm v_k^H\bm h_1^k|^2\sigma_s^2+\sigma_w^2}{|\bm v_k^H\bm h_0|^2\sigma_s^2+\sigma_w^2}
   		\\&\quad ~{=} \frac{(|\bm v_k^H\bm h_1^k|^2{-}|\bm v_k^H\bm h_0|^2)\sigma_s^2}{|\bm v_k^H\bm h_0|^2\sigma_s^2+\sigma_w^2}+1\ge 1, \\ 
   		& \bar x_k{=}\frac{|\bm { v}_k^H\bm {\bar h}_1^k|^2\sigma_s^2+\sigma_w^2}{\sigma_w^2}{=}\frac{|\bm { v}_k^H\bm {\bar h}_1^k|^2\sigma_s^2}{\sigma_w^2}+1\ge 1.\\ 
   	\end{aligned} \right.
   \end{equation}

Thus, $C_{3-2}$ can be written as 
\begin{equation} \label{wdl2} 	
	f(x_k)=\ln x_k+\frac{1}{x_k}\ge f^{\min},\\
\end{equation} 
where $f^{\min}=\frac{D^{\min}}{N}+1$.

\begin{figure}[t]
	\centerline{\includegraphics[width=3.2in]{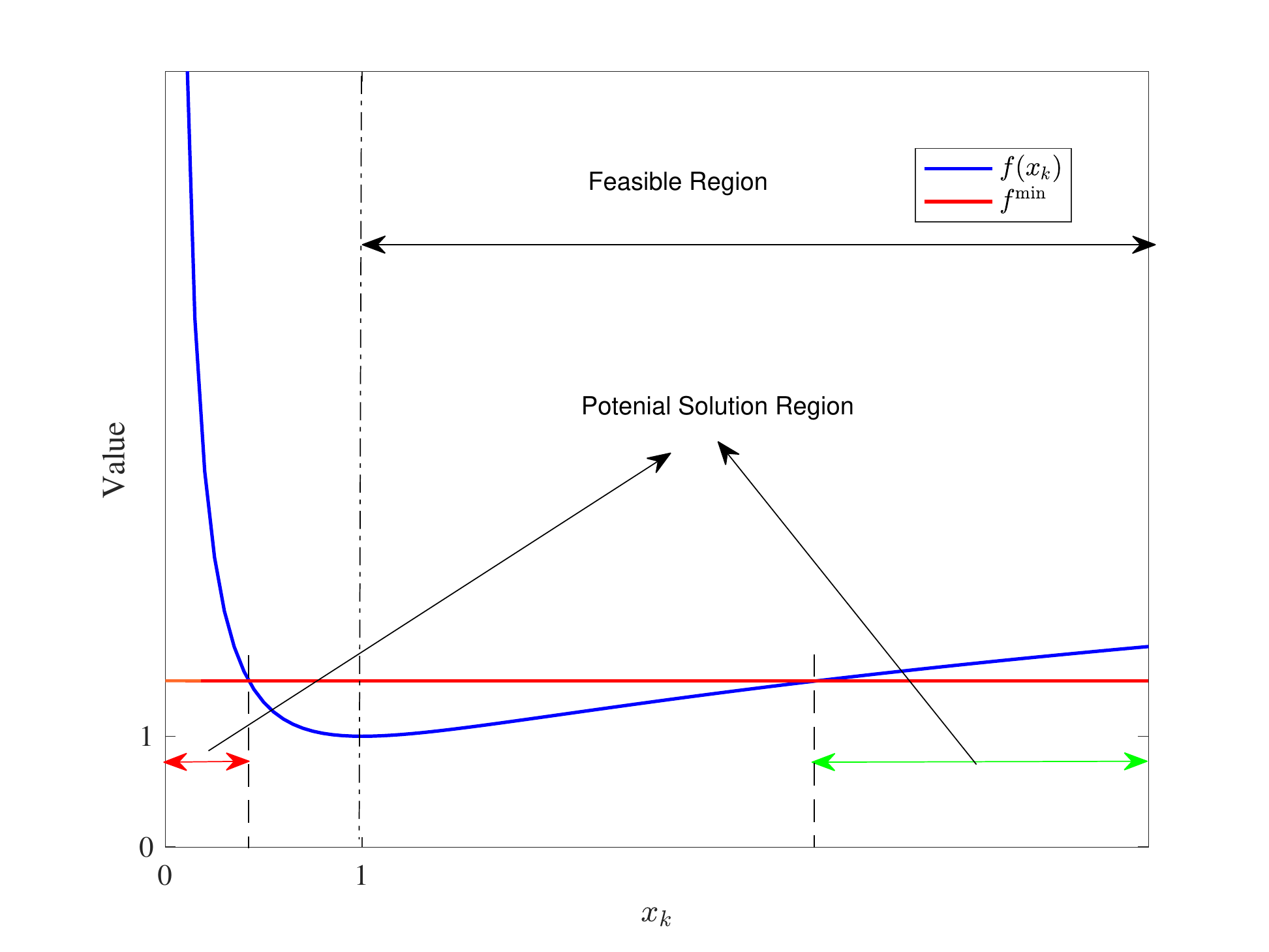}}
	\caption{The schematic diagram for $f(x_k)$.}
	\label{fig2}
\end{figure} 

As shown in the Fig. \ref{fig2},  there are two potential solution regions for  $x_k$ satisfying (\ref{wdl2}). According to the limitation of the feasible region in (\ref{15}), we can round off a solution region. However, it is still challenging to solve this inequality due to its non-convexity. To this end, taking (\ref{wdl2}) as an equivalent transformation, we have 
\begin{equation} \label{wdl4}
	\begin{aligned}
		&(\ln x_k{-}f^{\min})\exp (\ln x_k{-}f^{\min})\ge {-}\exp ({-}f^{\min}).\\
	\end{aligned}
\end{equation}
Since ${-}\exp ({-}f^{\min}) \in [{-}\exp ({-}1),0 )$, based on Lambert $W$ function \cite{ChapeauMonirtsp}, we have
\begin{equation} \label{wdl5} 
	\left\{ \begin{aligned}
		& \ln x_k{-}f^{\min}\le W_{{-}1}({-}\exp(-f^{\min})), 0{<}x_k{<}1, (\text{Neglect})\\
		& \ln x_k{-}f^{\min}\ge W_0(-\exp(-f^{\min})), x_k\ge 1.\\
	\end{aligned} \right.
\end{equation}
Thus, we can obtain the solutions of $x_k$, i.e.,
\begin{equation} \label{wdl6} 
	\left\{ \begin{aligned}	
		& x_k\le \bar F(f^{\min}), 0{<}x_k{<}1, (\text{Neglect})\\
		& x_k\ge F(f^{\min}), x_k\ge 1,\\
	\end{aligned} \right.
\end{equation}
where $\bar F(x)=\exp(W_{-1}(-\exp(-x))+x)$ and $F(x)=\exp(W_{0}(-\exp(-x))+x)$.

Similarly, $C_{4-2}$ can be rewritten as 
\begin{equation} \label{wdl7} 
	\begin{aligned}	
		& \bar x_k\ge F(g^{\min}),\\
	\end{aligned} 
\end{equation}
where $g^{\min}\triangleq\frac{E^{\min}}{N}+1$. This completes the proof.

	\section{Proof of  Theorem 2}	

Based on (\ref{14}) and (\ref{15}), $\bar C_{3-2}$ can be rewritten as
   \begin{equation} \label{15a} 	
\ln x_k+\frac{1}{x_k}-(\ln \bar x_k+\frac{1}{\bar x_k})\ge 0.\\
\end{equation}
According to its monotonicity, we can equivalently convert $f(x_k)-f(\bar x_k)\ge 0$ into $x_k\ge \bar x_k$ and $x_k<0< \bar x_k$ (which is neglected according to the  feasible region). That is to say, $\bar C_{3-2}$ can be transformed by the following form, i.e., 
   \begin{equation} \label{16}
   	\frac{|\bm {\hat v}_k^H\bm h_1^k|^2{-}|\bm {\hat v}_k^H\bm h_0|^2}{|\bm {\hat v}_k^H\bm h_0|^2\sigma_s^2+\sigma_w^2}\ge\frac{|\bm {\hat v}_k^H\bm {\bar h}_1^k|^2}{\sigma_w^2},\\
   \end{equation} 
   or equivalently
   \begin{equation} \label{16b}
   	\frac{{|\bm {\hat v}_k^H\bm h_1^k|^2{-}|\bm {\hat v}_k^H\bm h_0|^2}{-}|\bm { v}_k^H\bm {\bar h}_1^k|^2}{|\bm {\hat v}_k^H\bm h_0|^2|\bm {\hat v}_k^H\bm {\bar h}_1^k|^2}\ge \gamma.\\	
   \end{equation} 
This completes the proof.

\section{Proof of Theorems 3}

According to (\ref{s3}), we can obtain
\begin{equation} \label{s6}
	\begin{aligned}
		\ln  \frac{\delta_1^k}{\delta_0}+\frac{\delta_0}{\delta_1^k}\ge \ln \frac{\bar \delta_1^k}{\bar \delta_0}+\frac{\bar \delta_0}{\bar \delta_1^k}. \\	
	\end{aligned}
\end{equation}
Based on (\ref{15a}) and (\ref{16}), (\ref{s6}) can be transformed into
\begin{equation} \label{s6a}
	\begin{aligned}
	\frac{\sigma_s^2|h_1^k|^2+\sigma_w^2}{\sigma_s^2|h_0|^2+\sigma_w^2}\ge 	\frac{\sigma_s^2|\bar h_1^k|^2+\sigma_w^2}{\sigma_s^2|\bar h_0|^2+\sigma_w^2},\\	
	\end{aligned}
\end{equation}
or equivalently
\begin{equation} \label{s7}
	\begin{aligned}
		&\frac{|h_{\text{SR}}+ h^k_{\text{STR}}|^2-|h_{\text{SR}}|^2}{\sigma_s^2|h_{\text{SR}}|^2+\sigma_w^2}\ge \frac{|h^k_{\text{STR}}|^2}{\sigma_w^2}\\	
		\Leftrightarrow &(|h_{\text{SR}}+ h_k^{\text{STR}}|^2-|h_{\text{SR}}|^2-|h^k_{\text{STR}}|^2)\sigma_w^2\\
		&\quad \quad \quad \ge\sigma_s^2|h^{\text{SR}}|^2|h^k_{\text{STR}}|^2\\
		\Leftrightarrow &\frac{(h_{\text{SR}})^Hh^k_{\text{STR}}+h_{\text{SR}}	(h^k_{\text{STR}})^H}{|h_{\text{SR}}|^2|h^k_{\text{STR}}|^2}\ge \gamma.\\	
	\end{aligned}
\end{equation}
Besides, according to Appendix A, (\ref{s4}) can be transformed into 
\begin{align} \label{s7a}
	\gamma \ge \frac{F(g^{\min})-1}{|h^k_{\text{STR}}|^2}.
\end{align}
Thus, the CI region with respect to $\gamma$ can be expressed as
		\begin{equation}
	\begin{aligned} \label{s7b}
		& \frac{F(g^{\min})-1}{|h^k_{\text{STR}}|^2}\le \gamma \le \frac{(h_{\text{SR}})^Hh^k_{\text{STR}}+h_{\text{SR}}	(h^k_{\text{STR}})^H}{|h_{\text{SR}}|^2|h^k_{\text{STR}}|^2}.\\	
	\end{aligned}
\end{equation}
This completes the proof.

\section{Proof of Theorems 4}
According to the definition of vectorial angle, we have
	\begin{equation} \label{a1}
	\begin{aligned}
	\cos \left\langle h_{\text{SR}}, h^k_{\text{STR}}\right\rangle=\frac{|\left\langle h_{\text{SR}}, h^k_{\text{STR}}\right\rangle|}{|h_{\text{SR}}||h^k_{\text{STR}}|}=\frac{(h_{\text{SR}})^Hh^k_{\text{STR}}}{|h_{\text{SR}}||h^k_{\text{STR}}|}.\\
	\end{aligned}
\end{equation}
Similarly, we can obtain	
	\begin{equation} \label{a2}
	\begin{aligned}
	\cos \left\langle h^k_{\text{STR}}, h_{\text{SR}}\right\rangle =\frac{(h^k_{\text{STR}})^Hh_{\text{SR}}}{|h_{\text{SR}}||h^k_{\text{STR}}|}.\\
	\end{aligned}
\end{equation}
According to (\ref{s7}), we have 
	\begin{equation} \label{a3}
	\begin{aligned}
       \cos \left\langle h_{\text{SR}}, h^k_{\text{STR}}\right\rangle +\cos \left\langle h^k_{\text{STR}}, h_{\text{SR}}\right\rangle \ge \gamma {|h_{\text{SR}}||h^k_{\text{STR}}|}.\\
	\end{aligned}
\end{equation}
Due to the mutuality of angles, we have
	\begin{equation} \label{a4}
	\begin{aligned}
		\theta_k=\angle  \left\langle h_{\text{SR}}, h^k_{\text{STR}}\right\rangle \le \arccos \frac{1}{2}\gamma{|h_{\text{SR}}||h^k_{\text{STR}}|}.\\
	\end{aligned}
\end{equation}
Further, it does not make sense when $\theta_k>\frac{\pi}{2}$ holds, so we have 
	\begin{equation} \label{a5}
	\begin{aligned}
		\theta_k& \le\min  \left\{\frac{\pi}{2}, \arccos \frac{1}{2}\gamma{|h_{\text{SR}}||h^k_{\text{STR}}|}\right\},
	\end{aligned}
\end{equation}
where $\gamma\ge \frac{F(g^{\min})-1}{|h^k_{\text{STR}}|^2}$.
This completes the proof.

\section{Problem Formulation and Solution}

		   \begin{itemize}
		   	\item \textit{Problem Formulation}
		   	
		   	By removing the CI constraint and regarding the DL as interference,  the original optimization problem  is transformed into one for maximizing the  signal-to-interference-plus-noise ratio (SINR) under scheme I, which is expressed as  
		   		\begin{equation} \label{ac1} 
		   		\begin{aligned}	
		   			& \underset{\bm {\tilde v}_k, \tilde \beta_k}{\mathop{\max}}\,  \sum\limits_{k=1}^K \tilde \beta_k \frac{|\bm {\tilde v}_k^H\bm h_{\text{STR}}^k|^2 \sigma_s^2}{|\bm {\tilde v}_k^H \bm h_{\text{SR}}|^2\sigma_s^2+\sigma_w^2} \\ 
		   			&\text{s.t.}~
		   			{\tilde {C}_{1}}:\sum\limits_{k=1}^{K}{\tilde \beta_k}=1, \tilde \beta_k\in \{0,1\},\\ 
		   			&\quad ~~{\tilde {C}_{2}}:||\bm {\tilde v}_k||^2=1,\forall k,\\ 
		   			&\quad ~~{\tilde {C}_{3}}: \sum\limits_{k=1}^K \tilde \beta_k \gamma{|\bm {\tilde v}_k^H\bm h_{\text{STR}}^k|^2}\ge  F(g^{\min})-1. \\
		   		\end{aligned}
		   	\end{equation}  
	   	
	   	\item \textit{Algorithm Design}
	   	
	   	It is obvious that the minimum-mean-square-error (MMSE) approach can be used to obtain the optimal receive beamforming vector $\bm {\tilde v}_k^*$ to problem (\ref{ac1}) with fixed tag selection factor $\tilde \beta_k$. And then, optimal $\tilde \beta_k^*$ can be solved by using the same greedy algorithm. In particular, the MMSE-based receive beamforming is written as 
	   \begin{equation} \label{ac2} 
	   		\begin{array}{l}	
               \!\!\bm {\tilde v}_k^*{=}\dfrac{\left( \bm h_{\text{SR}}(\bm h_{\text{SR}})^H+\bm h_{\text{STR}}^k(\bm h_{\text{STR}}^k)^H+\frac{\sigma_w^2}{\sigma_s^2}\bm I_N\right)^{-1}\bm h_{\text{STR}}^k }{\left\|\left( \bm h_{\text{SR}}(\bm h_{\text{SR}})^H+\bm h_{\text{STR}}^k(\bm h_{\text{STR}}^k)^H+\frac{\sigma_w^2}{\sigma_s^2}\bm I_N\right)^{-1}\bm h_{\text{STR}}^k \right\|}.
	   		\end{array}
	   	\end{equation}  
		   \end{itemize} 

\end{appendices}

\end{document}